\documentclass[fleqn,usenatbib]{mnras}

\usepackage{newtxtext,newtxmath}

\usepackage[T1]{fontenc}

\DeclareRobustCommand{\VAN}[3]{#2}
\let\VANthebibliography\thebibliography
\def\thebibliography{\DeclareRobustCommand{\VAN}[3]{##3}\VANthebibliography}

\usepackage{graphicx}	
\usepackage{amsmath}	
\usepackage{natbib}

\usepackage{lineno}

\usepackage[version=4]{mhchem}

\title[Equifinality of Venus-like \ce{CO2} Atmospheres]{Equifinality of Venus-like \ce{CO2} Atmospheres}

\author[T. Constantinou et al.]{
Tereza Constantinou,$^{1}$\thanks{E-mail: tc496@cam.ac.uk}
Oliver Shorttle,$^{1,2}$
and Harrison Nicholls$^{1}$
\\
$^{1}$Institute of Astronomy, University of Cambridge, Madingley Road, Cambridge, CB3 0HA, UK\\
$^{2}$Department of Earth Sciences, University of Cambridge, Downing Street, Cambridge CB2 3EQ, UK\\
}

\date{Accepted XXX. Received YYY; in original form ZZZ}

\pubyear{\the\year{}}

\begin{document}
\label{firstpage}
\pagerange{\pageref{firstpage}--\pageref{lastpage}}
\maketitle

\begin{abstract}
While Earth locks much of its carbon in its crust as carbonates, Venus retains a comparable carbon inventory almost entirely in its atmosphere as \ce{CO2}. On Earth, the geological carbon cycle that has produced this vast crustal carbonate inventory is regulated by biology, liquid water, and plate tectonics, which together have stabilised climate over geological timescales. Venus presently lacks all these processes. We test whether Venus's massive \ce{CO2} atmosphere is diagnostic of a specific evolutionary pathway by quantifying three routes: primary magma-ocean outgassing, secondary volcanic degassing in a stagnant-lid regime, and remobilisation of crustal carbonates after climate destabilisation. Using a coupled climate--weathering framework, we find that a past habitable Venus could have stored $\sim$20~bar of \ce{CO2} as crustal carbonates. Following transition to runaway conditions, crustal heating releases this reservoir over tens of Myr. In stagnant-lid secondary-degassing models with a MORB-like mantle, outgassing reaches only $\sim$25~bar \ce{CO2}, limited by progressive mantle volatile depletion. However, Venus-like inventories can be achieved through: (i) magmatic carbon enrichment, (ii) increased magmatic delivery to the surface (high extrusion or melt production), and (iii) the recycling of undegassed carbon back into the planet's interior.  Primary magma-ocean outgassing can generate $>10^2$~bar \ce{CO2}, but the retained fraction after early escape remains uncertain.  Ultimately, a Venus-like massive \ce{CO2} atmosphere is an equifinal outcome and does not uniquely diagnose a temperate past.
\end{abstract}




\section{Introduction}
Comparative planetology provides a framework for understanding the processes that shape rocky planets across the Solar System and beyond. By examining commonalities, we can identify general principles of planetary evolution; by studying differences, we can identify phenomena that have been particular and contingent to one planet's evolution \citep[e.g.,][]{constantinou2025comparative}.  Of primary interest among these seemingly contingent and particular outcomes of planetary evolution is their arriving at a habitable state.

Within the solar system, a striking comparative case study for habitability lies in Earth and Venus: two neighbouring terrestrial planets that began with comparable bulk compositions \citep{morbidelli2012building,izidoro2022origin, peslier2022water}, but which now host distinct atmospheres \citep{seiff1985models}, surface environments \citep{ivanov2011map}, and perhaps interior compositions \citep{constantinou2024dry}. Earth’s atmosphere is temperate and \ce{N2}-dominated, with only a small fraction of its total carbon budget present as atmospheric \ce{CO2}; most of Earth's surficial carbon is stored in crustal carbonates and sediments, accumulated over geologic time by plate tectonics and aqueous chemistry \citep{hayes2006carbon,walton2024epsl}. Venus, by contrast, has a dense \ce{CO2}-rich atmosphere, with a surface pressure of 92 bars --- of which approximately 88 bars is \ce{CO2} \citep{seiff1985models}.

Despite the marked differences between modern Earth and Venus, it has been suggested that in the past they may have been more similar. Elevated atmospheric D/H ratios \citep{donhue} and climate modelling \citep{Way2016, Way2020} have been used to argue that Venus may have once hosted surface water and maintained temperate conditions. Understanding Venus's climate history is key to providing empirical constraints on the habitable zone concept \citep{kasting1993habitable}. The planets we are best able to discover and characterise are at short orbital period, lying at the inner edge of their system's liquid water habitable zone. If former temperate worlds can be found among this short period exoplanet population, we will be able to map the inner edge of the habitable zone and start constraining the occurrence of habitability among rocky planets \citep{jordan2025sciadv}. 

To achieve this, we need to develop diagnostics that allow us to distinguish uninhabitable planets which were formerly Earth-like, from those born Venus-like. Nitrogen is chemically inert and remains stable in the atmosphere over long timescales \citep{marty2012origins}, so atmospheric \ce{N2} can be used to decipher the outgassing history of a planet \citep{weller2023venus}. However, \ce{N2} is spectrally inactive in the visible and infrared, making it difficult to observe in exoplanet atmospheres. In contrast, \ce{CO2} is spectrally active, and can be probed directly in Earth-like exoplanet atmospheres \citep{gialluca2021characterizing, ostberg2023reading}. We therefore focus our analysis on Venus's and Earth's respective \ce{CO2} reservoirs. 

In principle, a massive \ce{CO2} atmosphere like Venus's could arise through three distinct mechanisms: (i) retention of a primary atmosphere outgassed during magma ocean crystallisation \citep{sossi_redox_2020, nicholls2024magma}, (ii) gradual accumulation through sustained secondary volcanic degassing into a persistently dry surface environment \citep{liggins2022growth, gaillard2021diverse}, or (iii) the breakdown of crustal carbonates following the climatic runaway of a formerly temperate world \citep{Way2020}.

\subsection{Earth-Venus geodynamics and volatile cycles}
\begin{figure}
    \centering
    \includegraphics[width=\columnwidth]{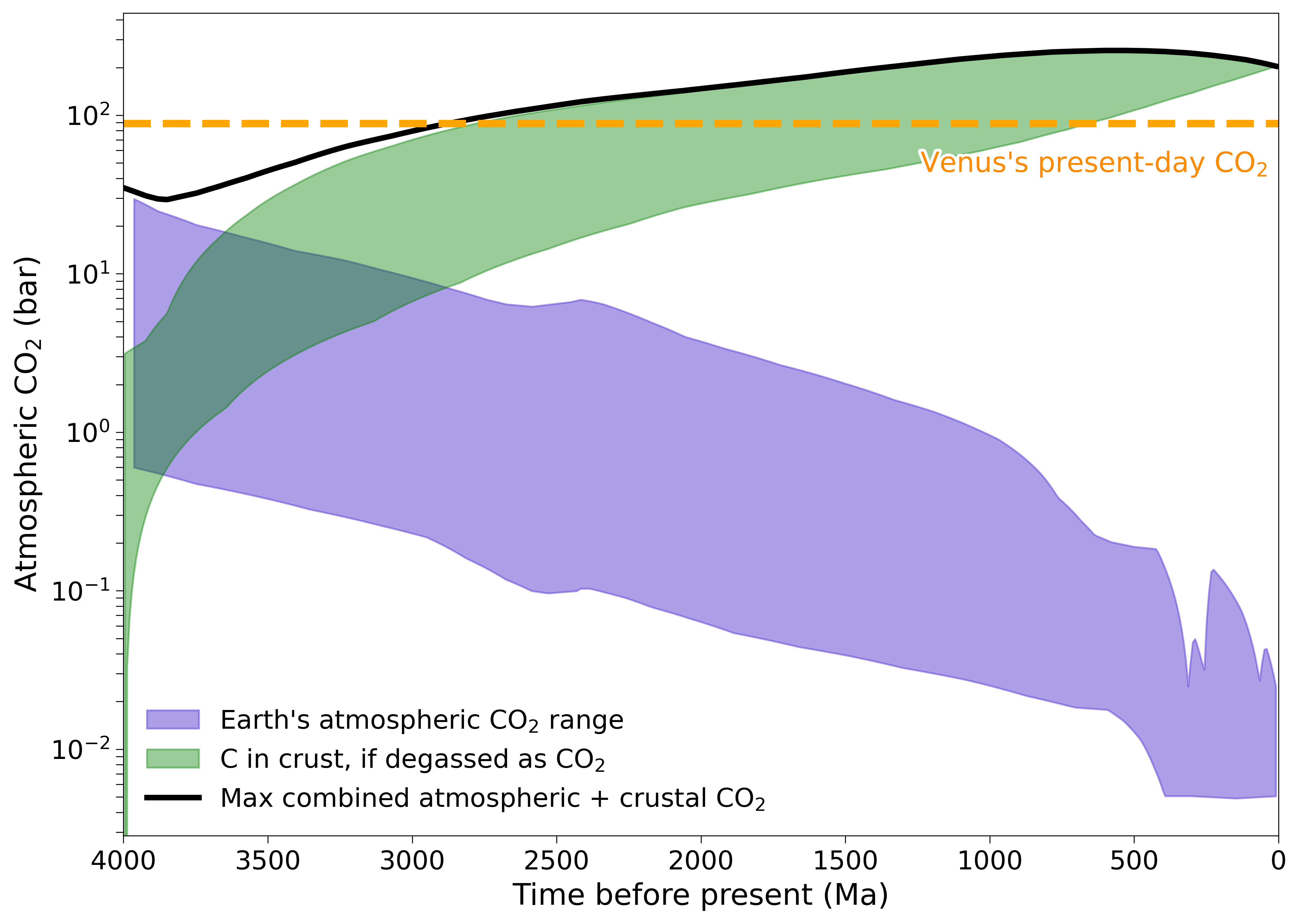}
    \caption{
    \textbf{The Earth's carbon reservoirs in the atmosphere and continental crust modelled over geological time.} The atmospheric \ce{CO2} concentration (purple) shows the 95\% confidence interval from a carbonate-silicate cycle model \citep{krissansen2018constraining, catling2020archean}. The carbon content of the continental crust is plotted as the equivalent atmospheric \ce{CO2} that would result if all crustal carbon were degassed as \ce{CO2} gas (total crustal carbon from \citet{hayes2006carbon}, and relative changes through time from \citet{ronov1964general, veizer2003evolution}). For comparison, the present-day atmospheric \ce{CO2} of Venus is shown as an orange dashed line.}
    \label{fig:all_C}
\end{figure}

The atmospheric \ce{CO2} reservoir of Venus is comparable in mass to Earth’s entire crustal carbon inventory \citep[Figure~\ref{fig:all_C};][]{lecuyer2000epsl,hayes2006carbon}.  This is suggestive of Venus having once perhaps shared a similar, water-mediated, atmosphere-surface volatile cycle to the Earth. 

Earth's carbon cycle begins with partial melting of the mantle, which produces basaltic crust and releases volatiles such as \ce{CO2} to the atmosphere. This basalt may be further processed to produce continental crust, which tectonics crumples to form mountain ranges.  Mountain ranges are sites of extensive rainfall focusing silicate weathering onto Earth's continental mountain ranges and basaltic seafloor to, ultimately, sequester carbon into carbonate minerals \citep{walker1981negative,coogan2026well}.  These carbonates may then be stored for hundreds of millions of years on the continents \citep{walton2024epsl}, or returned into the mantle via subduction.

In contrast, on present-day Venus this global-scale carbon recycling mechanism is absent \citep[][and references therein]{rolf2022dynamics}. While not geologically inert, Venus's tectonic regime remains debated. One proposal is a `plutonic-squishy lid’, where high rates of intrusive magmatism create a mosaic of small, mobile blocks separated by thermally weakened boundaries \citep{lourenco2020plutonic}. Alternatively, subsequent modelling has suggested that incorporating a more realistic crustal rheology, accounting for weak plagioclase, facilitates a distinct `deformable episodic lid’ regime characterised by high, continuous surface mobility and widespread deformation between larger-scale lithospheric overturns \citep{tian2023tectonics}. Crucially, both of these proposed non-plate tectonic regimes are dominated by vertical material recycling through delamination and ductile dripping, rather than the large-scale, coherent downwelling of modern subduction that is especially efficient for long-term sequestration of surface volatiles \citep[e.g., ][]{chen2023carbonate}.

The view of an early Venus having plate tectonics is supported by recent efforts to reconcile the planet's atmospheric \ce{N2} inventory with its degassing history. \citet{weller2023venus} show through long-term evolution models that a continuous stagnant-lid history is incapable of outgassing sufficient \ce{N2} to account for its modern atmospheric abundance. Conversely, a continuous history of plate tectonics would place substantially too much \ce{N2} in Venus's atmosphere. \citet{weller2023venus} therefore conclude Venus's modern atmosphere results from an early period of plate tectonic-like activity lasting for at least a billion years, followed by a transition to the current stagnant-lid-like state. We note, however, that this interpretation remains challenging: sustaining plate tectonics in a hotter mantle may be dynamically difficult \citep[e.g.,][]{van2008plate}, and evidence from early Earth suggests Archean tectonic regimes may have differed from modern plate tectonics \citep[as informed from paleomagnetic evidence in][]{brenner2026paleomagnetic}.

This view of Venus as having a previously Earth-like past with oceans is potentially at odds with climate models. When initiated in a hot, ocean-free state, Venus climate models have shown that the planet may never have cooled sufficiently to allow water condensation from the atmosphere \citep{Turbet2021}. Recent constraints on Venus's mantle water contents support this conclusion, finding a dry Venusian interior \citep{constantinou2024dry}, inconsistent with a past hydrological cycle like that of the Earth. In this hotter, ocean-free scenario, crustal carbonates could not have formed or been stable at the surface \citep[as they are presently not stable;][]{treiman2012mineral} ---  and the present-day atmosphere would instead be the product of continuous volcanic degassing over time \citep[e.g.,][]{gaillard2014theoretical,liggins2022growth, gaillard2021diverse}.

Alternatively, if Venus retained even a fraction of a massive primary atmosphere outgassed during its magma ocean stage \citep{nicholls2024magma}, the modern inventory could reflect this early volatile budget rather than sustained interior processing. The question is whether these three evolutionary scenarios, retention of primary magma ocean degassing; secondary volcanic degassing; or crustal decarbonation, lead to distinct atmospheric outcomes that can be observed in planets beyond the Solar System.

\subsection{Structure of the paper}
Here, we test the distinctness of these three end-member scenarios for Venus's evolution, in relation to their eventual atmospheres: one scenario is a transiently temperate rocky world turned `dead Earth'; another is a `born Venus'; and a third is a Venus that retained its primary magma ocean-outgassed atmosphere. We perform this test using coupled atmosphere–interior models. In Section~\ref{sec:build_second}, we estimate the maximum atmospheric \ce{CO2} that can be built up through continuous degassing under stagnant-lid conditions. In Section~\ref{sec:c_release}, we calculate the carbon that could be sequestered during a temperate phase, later remobilised following climate destabilisation. By comparing the outcomes of these models to Venus’s present-day atmosphere in Section~\ref{sec:secondary}, we assess whether its dense \ce{CO2} inventory is more consistent with primary outgassing (Section~\ref{sec:discuss_primary}), gradual secondary degassing (Section~\ref{sec:discuss_secondary}), or a catastrophic crustal remobilisation (Section~\ref{sec:discuss_hab}). Section~\ref{sec:discuss_exo} discusses the observational challenge of distinguishing between these evolutionary scenarios. Finally, Section~\ref{sec:concl} concludes our findings and their implications for constraining the climatic limits of Earth-like worlds and interpreting \ce{CO2}-rich atmospheres on terrestrial exoplanets.

\section{Methods}
\subsection{Building a secondary atmosphere}\label{sec:build_second}
To simulate the growth of a planet's secondary atmosphere, we use the EVolve model \citep{liggins2022growth}, which links mantle melting to atmospheric composition via volcanic outgassing. EVolve calculates volatile partitioning during partial melting according to the batch melting equation
\begin{align}\label{eq:batch}
    X_{i,\mathrm{melt}} = \frac{X_{i,\mathrm{mantle}}}{D_i + F(1 - D_i)},
\end{align}
where $X_{i,\mathrm{melt}}$ and $X_{i,\mathrm{mantle}}$ are the concentrations of species $i$ in the melt and bulk-mantle, respectively, $F$ is the mantle melt mass-fraction (fixed at 0.1), and $D_i$ is the partition coefficient \citep[see Table 2 in][]{liggins2022growth}. Because $D_i < 1$ for most volatile elements \citep[e.g.,][]{rosenthal2015experimental,aubaud2004hydrogen}, partial melts of the mantle form enriched in these elements compared to the planet's interior.  One exception to this is when carbon is in reduced form as graphite in the mantle, in which case partitioning between solid graphite and the silicate melt will take place \citep[e.g., as parametrised by][]{eguchi2018co2}.  However, the presence of reduced carbon phases decreases the efficiency of carbon transport to the surface. Because we seek to determine the maximum possible mass of a volcanic atmosphere, we neglect graphite saturation. Our results therefore represent an optimistic upper bound on the volcanic carbon inventory \citep{bower_diversity_2025}.

The melt compositions calculated using Equation \eqref{eq:batch}, along with how oxidising the mantle is ($f$\ce{O2}, expressed relative to the iron--w{\"u}stite buffer as $\Delta$IW), serve as input to the EVolve model \citep{liggins2020can}, which computes the mass, composition, and speciation of the volcanic gas mixture at the current surface pressure. The speciation is computed using the equilibrium constant and mass balance method, and comprises 10 C–O–H–S–N species (\ce{H2O}, \ce{H2}, \ce{O2}, \ce{CO2}, \ce{CO}, \ce{CH4}, \ce{S2}, \ce{SO2}, \ce{H2S}, \ce{N2}) by coupling homogeneous gas-phase equilibria,
\begin{align}
    \ce{H2} + \tfrac{1}{2}\ce{O2} &\rightleftharpoons \ce{H2O}, \\
    \ce{CO} + \tfrac{1}{2}\ce{O2} &\rightleftharpoons \ce{CO2}, \\
    \ce{CH4} + 2\ce{O2} &\rightleftharpoons \ce{CO2} + 2\ce{H2O}, \\
    \ce{H2S} + \tfrac{1}{2}\ce{O2} &\rightleftharpoons \tfrac{1}{2}\ce{S_2} + \ce{H2O}, \\
    \tfrac{1}{2}\ce{S2} + \ce{O2} &\rightleftharpoons \ce{SO2},
\end{align}
with heterogeneous melt-gas equilibria suitable for basaltic melts. Solubility laws are adopted from \cite{burgisser2015simulating} (\ce{H2O}, \ce{H2}), \cite{eguchi2018co2} (\ce{CO2}), \cite{armstrong2015speciation} (\ce{CO}), \cite{ardia2013solubility} (\ce{CH4}), \cite{libourel2003nitrogen} (\ce{N2}), and \cite{o2021thermodynamic} (sulfide capacity). \ce{CO} and \ce{CH4} are treated as insoluble if $f\ce{O2} > \mathrm{IW}+1$. 

The oxidation state of the erupting system is governed by the ferric-ferrous iron equilibrium in the melt \citep{kress1991compressibility},
\begin{equation}
\ce{Fe2O3(melt)} \rightleftharpoons 2\ce{FeO(melt)} + \tfrac{1}{2}\ce{O2(gas)},
\end{equation}
which imposes equilibrium between the gas and silicate phases. While the system's $f\mathrm{O}_2$ evolves during decompression, the melt's iron inventory provides a substantial redox buffer, with the initial redox state prescribed as a fixed offset from a mineral buffer (here $\Delta$IW), such that
\begin{equation}
\log_{10} f\mathrm{O}_2 = \log_{10} f\mathrm{O}_{2,\mathrm{buffer}}(T,P) + \Delta\mathrm{IW}.
\end{equation}
At each timestep, this buffer-relative \(f\mathrm{O}_2\) is passed to EVo, which then solves melt--gas speciation self-consistently during decompression \citep{kress1991compressibility,burgisser2007redox}.

Outgassing is simulated by first finding volatile saturation (\(\sum_i P_i=P_{\mathrm{total}}\)) and then decompressing stepwise to the surface pressure at fixed T=1473\,K. We adopt this constant temperature from \citet{liggins2022growth}: it represents a plausible basaltic melt temperature and isolates the pressure/redox controls on speciation without introducing additional uncertainty from time-varying melt temperature.

We initialise the model with a volatile-rich mantle inventory representing conditions after magma ocean crystallisation: 450~ppm \ce{H2O} and 50~ppm \ce{CO2} \citep{elkins2008linked}. In the absence of direct constraints on mantle S and N abundances, we scale these to match Earth's depleted mid-ocean ridge basalt (MORB) source, yielding 54~ppm S \citep{ding2017fate} and 1~ppm N \citep{marty2003nitrogen}. On Earth, the MORB source is depleted by loss of volatiles to the crust, ocean, and atmosphere; on a hot Venus, the two former reservoirs may be absent, so a larger fraction of these elements could instead remain in the atmosphere. The configuration used here follows the standard configuration in \citet{liggins2022growth}.

We adopt a constant mantle melt production rate, $M_{\text{melt}} = 1\times10^{15}$~kg~yr$^{-1}$, appropriate for an Earth-sized planet \citep{ortenzi2020mantle}. The extrusive mass erupted to the surface is calculated by 
\begin{equation}\label{eqn:melt}
M_{\text{ext}} = r_{ei} M_{\text{melt}} \text{d}t, 
\end{equation}
where d$t$ is the time-step and $r_{ei}=0.1$ is the extrusive fraction. This results in a surface flux of $1\times10^{14}$~kg~yr$^{-1}$ ($\sim 27$~km$^3$~yr$^{-1}$), consistent with lower-bound estimates for Earth's mid-ocean ridges \citep{li2016quantifying}. We note that in this framework, the melt production rate serves only to scale the absolute time evolution of the system, by setting the rate at which the atmosphere accumulates.  The choice of melt flux does not alter the trajectory or steady-state composition of the outgassed volatiles, but does influence the final mass of the atmosphere.

Atmospheric composition is tracked dynamically, with gases released into an initial trace 0.01~bar \ce{N2} atmosphere. This choice defines a purely secondary-atmosphere experiment: we initialise with a negligible background atmosphere and follow subsequent atmospheric growth from volcanic outgassing alone. At each time-step, the combined atmosphere is updated: the new surface pressure  $P_{j}$ (Pa)  is determined by 
\begin{equation}
    P_{j} = P_{j-1} + \frac{M_{\text{ext}} W_{g} g}{4 \pi R_{p}^{2}},
    \label{eq:p_atmo}
\end{equation}
where $ P_{j-1}$ is the surface pressure from the preceding time-step (i.e., the pressure into which the volcanic gas is released, generated by the atmosphere already in place), $ W_{g}$ is the weight fraction of exsolved gas in the volcanic system, $g$ (m\,s$^{-2}$) is the surface gravity and $R_{p}$ (m) is the radius of the planet. Equilibrium chemical speciation of the atmosphere is computed using FastChem \citep{stock2022fastchem}, resolving 85 gas-phase species.

Atmospheric equilibrium chemistry is evaluated at the surface using a fixed surface temperature, with no coupling to climate. Volcanic degassing is computed separately with EVolve at fixed magmatic temperature (1473\,K), decompressed to the current surface pressure. In this framework, climate does not feed back on degassing; degassing is controlled by surface pressure together with melt composition, redox state, and volatile inventory. This decoupling allows straightforward comparison across mantle compositions.  All model planets are assumed to be Earth-sized (R$_p$=6.371$\times10^6$~m, g=9.81~m~s$^{-2}$). We set the rocky-planet mass parameter to $5.972\times10^{27}$~g and the mantle mass to $4.01\times10^{27}$~g ($\sim67$~wt\%), implying a non-mantle (core+crust) mass of $\sim1.96\times10^{27}$~g ($\sim33$~wt\%). EVolve does not explicitly resolve core radius/structure; core size enters only implicitly through this mass partition.

We consider two degassing scenarios. (1) \textit{No volatile recycling}: melt extracted from the mantle is partitioned into extrusive and intrusive components; the intrusive component plus the degassed residual of erupted melt are stored in a  crustal/lithospheric reservoir, so mantle mass and volatile concentrations decrease over time. (2) \textit{Volatile recycling}: no persistent crustal/lithospheric reservoir is formed; after degassing, the residual extrusive melt is mixed back into the mantle.

In recycling mode, for each volatile species $i$, EVolve updates mantle concentration by first removing the erupted melt from the mantle and then adding back the degassed residual melt. Here, $X_{i,\mathrm{mantle}}$, $X_{i,\mathrm{melt}}$, and $X_{i,\mathrm{res}}$ are concentrations of species $i$ (ppm) in the mantle, erupted melt before degassing, and degassed residual melt, respectively; $M_{\mathrm{mantle}}$, $M_{\mathrm{ext}}$, and $M_{\mathrm{res}}$ are the corresponding reservoir masses; $t$ is model time and $\mathrm{d}t$ is one model timestep
\begin{align}
&X_{i,\mathrm{mantle}}(t+\mathrm{d}t)=\\
&\frac{X_{i,\mathrm{mantle}}(t)\,M_{\mathrm{mantle}}(t)
- X_{i,\mathrm{melt}}(t)\,M_{\mathrm{ext}}(t)
+ X_{i,\mathrm{res}}(t)\,M_{\mathrm{res}}(t)}
{M_{\mathrm{mantle}}(t)-M_{\mathrm{ext}}(t)+M_{\mathrm{res}}(t)},
\end{align}
with
\begin{equation}
M_{\mathrm{res}}(t)=M_{\mathrm{ext}}(t)\,[1-W_g(t)],
\end{equation}
where $W_g$ is the exsolved gas mass fraction.

For both cases of volatile cycling regimes, we track the evolving atmospheric composition and pressure, focusing on the role of initial mantle carbon and hydrogen content (here set by mantle \ce{CO2} and \ce{H2O}) and magmatic $f$\ce{O2} state in setting the size and speciation of the resulting atmosphere.

\subsection{Carbon release from a destabilised crust} \label{sec:c_release}
We estimate the maximum mass of \ce{CO2} that silicate weathering could sequester into a temperate Venusian crust. This upper bound represents a strict physical limit, as it assumes all degassed \ce{CO2} is quantitatively removed from the atmosphere and permanently stored as carbonate. This calculation is independent of the volcanic models in Section~\ref{sec:build_second} in which mantle carbon availability and atmospheric pressure regulate outgassing.

The key quantity is the largest volcanic \ce{CO2} degassing flux that can be balanced by silicate weathering under a coupled climate--carbon equilibrium, such that \ce{CO2} does not accumulate in the atmosphere and trigger a runaway greenhouse. To identify this limit, we use the coupled climate--geochemical framework of \citet{graham2020thermodynamic}. For a given surface temperature $T_s$, the equilibrium atmospheric \ce{CO2} partial pressure is obtained by solving the planetary energy balance
\begin{equation}
    \frac{S_{\mathrm{eff}} (1 - a_p)}{4} = \mathrm{OLR}(T_s, p\text{CO}_2),
\end{equation}
where $S_{\mathrm{eff}}$ is the stellar insolation and $a_p = 0.3$ is the planetary albedo \citep[Earth's planetary albedo;][]{stephens2015albedo}. We consider two insolation states at Venus’s orbit: an early-Sun case with $S_{\mathrm{eff}}=0.75\,S_{0,\mathrm{Venus}}$ ($\approx 1.96\times10^{3}$~W~m$^{-2}$) and a present-day case with $S_{\mathrm{eff}}=1.0\,S_{0,\mathrm{Venus}}$ ($\approx 2.61\times10^{3}$~W~m$^{-2}$). Outgoing long-wave radiation (OLR) is parametrised using polynomial fits to convective-climate model output \citep{graham2020thermodynamic}; separate fits are used below and above 1~bar (Appendix~\ref{appendix:OLR}; coefficient matrices in Appendix Tables~\ref{tab:olr_low} and \ref{tab:olr_high}). This OLR parametrisation is calibrated to the moist greenhouse regime and does not reproduce the OLR modelled for cool, water-condensed atmospheres ($T_s \approx 288$~K); we therefore do not use this framework to model Earth directly.

The global silicate weathering flux $W_{\mathrm{sil}}$ is then computed using the Maher--Chamberlain (MAC) formulation \citep{maher2014hydrologic},
\begin{align}
    &W_{\mathrm{sil}} = \gamma A_{\mathrm{planet}} \cdot
    \frac{\alpha}{
    k_{\mathrm{eff}}^{-1}
    \!\!+ m A_s t_s
    + \frac{\alpha}{q_r \, \lambda (p\text{CO}_2)^n}
    },
\end{align}
where $A_{\mathrm{planet}}=4\pi R_{Venus}^2$ is the planetary surface area, and $\gamma = 0.3$ is the exposed land fraction \citep{Way2016,Way2020}.
The composite weathering-zone parameter is
\begin{align}
\alpha = L\,\phi\,\rho_{\mathrm{sf}}\,A_{\mathrm{s}}\,X_r\,\mu,
\end{align}
with $L$ flow-path length, $\phi$ porosity, $\rho_{\mathrm{sf}}$ mineral-mass to fluid-volume ratio, $A_{\mathrm{s}}$ specific reactive surface area, $X_r$ reactive mineral concentration, and $\mu$ the transit-time factor. 
The effective kinetic term is
\begin{align}
k_{\mathrm{eff}} = k_{\mathrm{eff,ref}}
\exp\!\left(\frac{T_s-T_{\mathrm{ref}}}{T_e}\right)
\left(\frac{p\mathrm{CO}_2}{p\mathrm{CO}_{2,\mathrm{ref}}}\right)^\beta,
\end{align}
where $k_{\mathrm{eff,ref}}$, $T_{\mathrm{ref}}$, $T_e$, $p\mathrm{CO}_{2,\mathrm{ref}}$, and $\beta$ are the reference kinetic constants. 
The $mA_{\mathrm{s}}t_s$ term is the supply/soil-age term, with $m$ molar mass and $t_s$ soil age. 
$q_r$ is global mean runoff \citep[capped by the energetic hydrological limit;][]{pierrehumbert2002hydrologic,ogormanschneider}, and $\lambda$ and $n$ are empirical constants defining the equilibrium solute concentration in runoff water: $C_{\mathrm{eq}}=\lambda (p\mathrm{CO}_2)^n$. Unless otherwise stated, parameter values follow \citet{graham2020thermodynamic}.

Here, we define a model threshold as the largest \ce{CO2} degassing rate for which a steady climate--carbon equilibrium exists within our adopted liquid-water/weathering regime ($T_s \le 373$~K). Above this model threshold, no solution in our framework simultaneously satisfies energy balance and carbon balance, so atmospheric \ce{CO2} accumulation is unavoidable. Note, we do not interpret 373~K as a universal runaway-greenhouse onset temperature; here it is an operational liquid-water/weathering cutoff. Classical runaway limits are largely derived from idealised 1-D pure-steam atmospheric structures, whereas more compositionally complex atmosphere--interior models can weaken or remove a unique OLR ceiling \citep{goldblatt2013low,selsis2023cool,boer2025runaway}.

We then model the propagation of surface heat into the crust following an abrupt transition from habitable to post-runaway conditions: surface temperature steps from $T_{\rm hab} = 373$~K (the adopted upper limit of liquid water stability) to $T_{\rm run} = 735$~K (Venus's present-day surface temperature; \citealt{seiff1985models}). We solve the one-dimensional heat equation in the crust
\begin{equation}
    \rho c_p \frac{\partial T}{\partial t} = k \frac{\partial^2 T}{\partial z^2} + A_r,
\end{equation}
where $\rho$ is crustal density, $c_p$ specific heat capacity, $k$ thermal conductivity, and $A_r$ volumetric radiogenic heat production. We adopt $\rho=2900$~kg~m$^{-3}$, $c_p=900$~J~kg$^{-1}$~K$^{-1}$, and $k=2.5$~W~m$^{-1}$~K$^{-1}$. The initial geotherm at $t=0$ includes background internal heating and is written as
\begin{equation}
    T(z, 0) = T_{\rm hab} + \frac{q}{k}\,z - \frac{A_r}{2k}\,z^2.
\end{equation}
where $q$ is surface heat flux. We use $q=65$~mW~m$^{-2}$ and $A_r=0.4~\mu$W~m$^{-3}$ \citep{turcotte2002geodynamics}. This $q$ choice is intended as a nominal value rather than a unique global estimate; Venus likely exhibits substantial spatial variability in heat flux \citep{smrekar2023earth, smrekar2025author}. We therefore also evaluate sensitivity to $q=50$--$80$~mW~m$^{-2}$. With $T_{\rm run}$ fixed at the surface and the base ($z = 50$~km) held at its initial value, the governing equation admits the analytical complementary-error-function (erfc) solution
\begin{equation}
    T(z,t) = T_{\rm run} + (T_{\rm hab} - T_{\rm run})\,
              \mathrm{erfc}\!\left(\frac{z}{2\sqrt{\kappa t}}\right),
\end{equation}
where $\kappa = k/(\rho c_p)$ is the thermal diffusivity. This semi-infinite approximation is valid for $t \ll H^2/\kappa \approx 80$~Myr ($H = 50$~km). The carbon release timescales of interest (17--78~Myr, discussed in Section~\ref{sec:cca}) lie comfortably within this regime; profiles shown at longer times in Figure~\ref{fig:decarb} are illustrative of the approach to the new steady state.

We compare the evolving geotherms against the pressure--temperature stability fields for nine decarbonation equilibria involving calcite (\ce{CaCO3}), magnesite (\ce{MgCO3}), and dolomite (\ce{CaMg(CO3)2}) in the presence of crustal silicates \citep{treiman2012mineral}, as detailed in Appendix~\ref{appendix:carbonates} and tabulated in Appendix Table~\ref{tab:decarb_reactions}. At each depth, equilibrium \ce{CO2} pressure is set by lithostatic pressure. We define $T_{\rm eq,min}(z)$ as the minimum equilibrium temperature across the nine reactions, i.e. the first reaction to become spontaneous at depth $z$. The habitable geotherm intersects this threshold at $z_{\rm cross}=23$~km (range 17--34~km for $q=50$--80~mW~m$^{-2}$; Figure~\ref{fig:decarb}a). Carbonates above $z_{\rm cross}$ are stable during the habitable phase and define the sequestered reservoir; after transition to post-runaway conditions, the geotherm lies in the instability field at all depths, so the stored column is thermodynamically destabilised. 

We account for latent heat of decarbonation using a Stefan correction with $M_{\rm CaCO_3}=0.100$~kg~mol$^{-1}$, $L_{\rm rxn}=100$~kJ~mol$^{-1}$, $X_{\rm carb}=7$\%, and $\Delta T=T_{\rm run}-T_{\rm hab}=362$~K. Liberated \ce{CO2} is then transported by buoyancy-driven Darcy flow and treated as conservative during ascent (no re-carbonation or in-transit trapping). We bracket permeability using two end-members: the \citet{manning1999permeability} continental-crust law, $\log_{10}k=-14-3.2\log_{10}(z/\mathrm{km})$ (upper release-rate bound), and a stagnant-lid proxy $k\times10^{-2}$ (lower bound). Darcy transit times are $<1\%$ of thermal-front arrival times at all depths, so heat diffusion is the rate-limiting process.

During the habitable phase, volcanic resurfacing buries the uppermost crust at a rate $v$~(m~yr$^{-1}$), where $v = \dot{V} / A_{\rm planet}$ and $\dot{V}$ is the volumetric resurfacing rate (km$^3$~yr$^{-1}$). Carbonates in the uppermost layer are continuously buried, and fresh surface rock is exposed for weathering. The time for burial to reach the crossover depth is $t_{\rm bury} = z_{\rm cross}/v$; for $t_{\rm hab} < t_{\rm bury}$ the crustal inventory grows linearly with time, while for $t_{\rm hab} \geq t_{\rm bury}$ it saturates at a ceiling value $C_{\rm ceil} = F_w \, t_{\rm bury}$, where $F_w$ is the maximum weathering flux computed above. At fixed total planetary age, we vary $t_{\rm hab}$ to explore the sequestration--release trade-off: longer habitable intervals increase carbonate sequestration (up to $C_{\rm ceil}$), but shorten the post-runaway interval available for thermal decarbonation and \ce{CO2} liberation before the present epoch. \ce{CO2} partial pressures are calculated from molar inventories via the hydrostatic relation $\Delta P = n M_{\rm CO_2} g / A_{\rm planet}$ (Appendix~\ref{appendix:bookkeeping}), where $n$ is the number of moles, $M_{\rm CO_2} = 0.044$~kg~mol$^{-1}$, $g = 8.87$~m~s$^{-2}$, and $A_{\rm planet}$ is the planetary surface area.

\section{Results}

\subsection{Secondary volcanic atmospheres}\label{sec:secondary}
\begin{figure*}
    \centering
    \includegraphics[width=\textwidth]{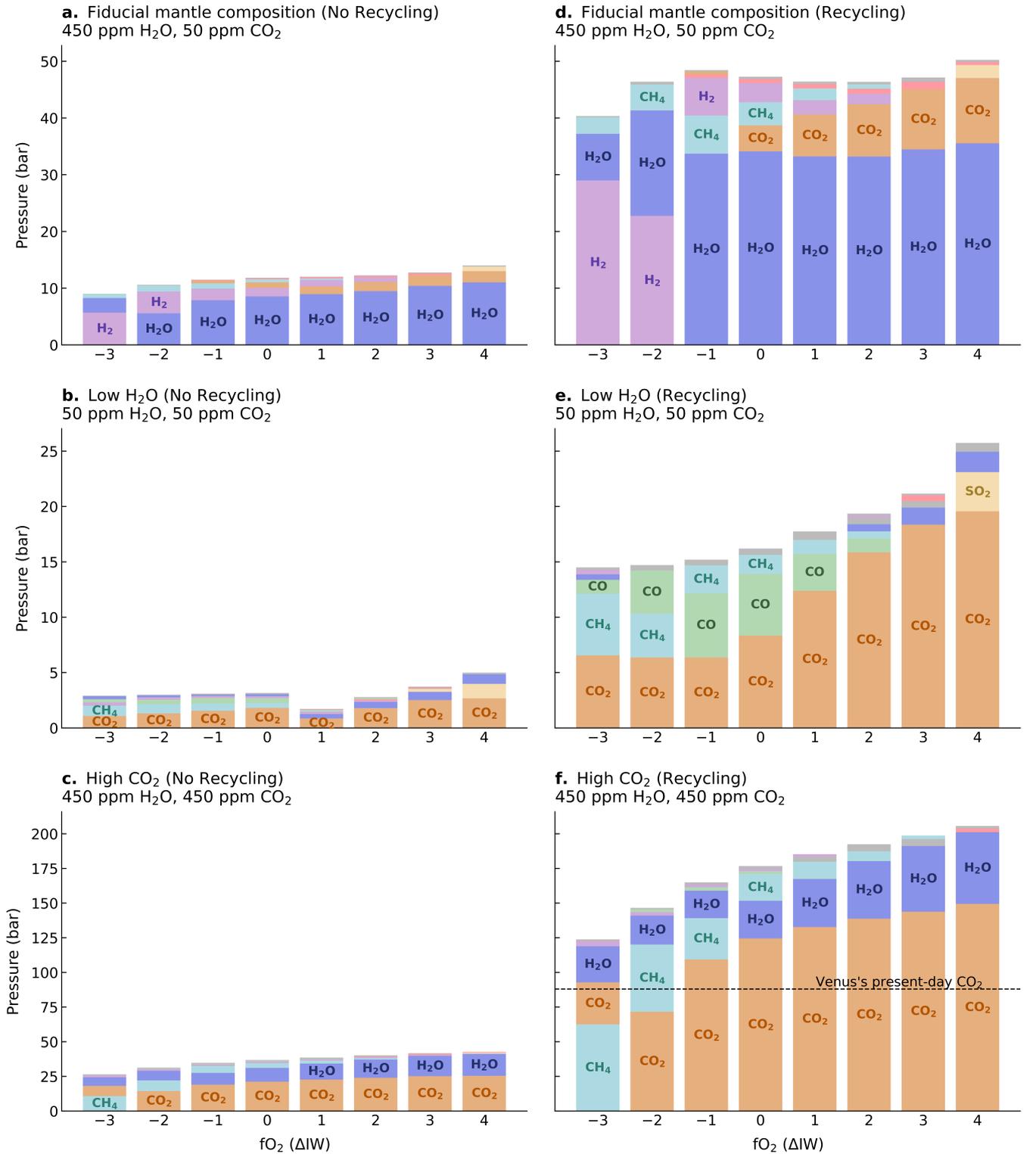}
    \caption{\textbf{Final atmospheric partial pressures after 4\,Gyr of volcanic degassing.} Stacked bars represent equilibrium partial pressures comparing the absence (left column; \textbf{a--c}) and presence (right column; \textbf{d--f}) of volatile recycling. Rows correspond to distinct mantle volatile inventories: \textbf{a, d,} Fiducial composition with 450\,ppm \ce{H2O} \citep{elkins2008linked} and 50\,ppm \ce{CO2} \citep{elkins2008linked}, containing 54\,ppm S and 1\,ppm N (scaled to Earth’s depleted MORB \citep{ding2017fate, marty2003nitrogen, le2017heterogeneity}). \textbf{b, e,} Reduced water content (50\,ppm \ce{H2O}). \textbf{c, f,} Increased \ce{CO2} content (450\,ppm \ce{CO2}). Cumulative bar heights indicate total atmospheric pressure. Atmospheric composition is shown as a function of mantle oxygen fugacity ($f\text{O}_2$), expressed relative to the iron--wustite buffer ($\Delta$IW). Major species are distinguished by colour and labelled; minor species ($<1\%$) are grouped in grey. }
    \label{fig:combined_composition}
\end{figure*}

To assess the capacity of secondary outgassing to generate Venus-like \ce{CO2} atmospheres, we model atmospheric evolution over 4~Gyr of volcanic degassing, varying the mantle carbon content, water content, and redox state. Figures~\ref{fig:combined_composition}a, b, and c show the resulting atmospheric compositions for Earth-sized planets under varying f\ce{O2} for three different mantle volatile inventories: a MORB-like fiducial case with 450~ppm \ce{H2O} and 50~ppm \ce{CO2} \citep{elkins2008linked}; a water-poor case with 50~ppm \ce{H2O}; and a carbon-rich case with 450~ppm \ce{CO2}, adopted as a heuristic Earth-informed sensitivity value within mantle-source carbon ranges \citep{hirschmann2009h,dasgupta2010deep}.

For a planet with no volatile recycling, intrusive melt and the degassed residual of erupted melt are stored in a long-lived crustal/lithospheric reservoir rather than returned to the mantle, even extremely oxidised and \ce{CO2}-enriched scenarios result in at most $\sim$25~bar equivalent of \ce{CO2} surface pressure (Figures \ref{fig:combined_composition}a, b, c). In this case, volatiles experience one-way transport out of the planetary interior. This is despite mantle \ce{CO2} enrichment of a factor of nine relative to Earth’s depleted MORB source. The inability to sustain further degassing arises because volatiles are progressively extracted from the mantle and sequestered into the growing crust, rather than into the atmosphere. In this case, there is no mechanism to return carbon to the mantle and prolong degassing.

\begin{figure*}
    \centering
    \includegraphics[width=0.96\textwidth]{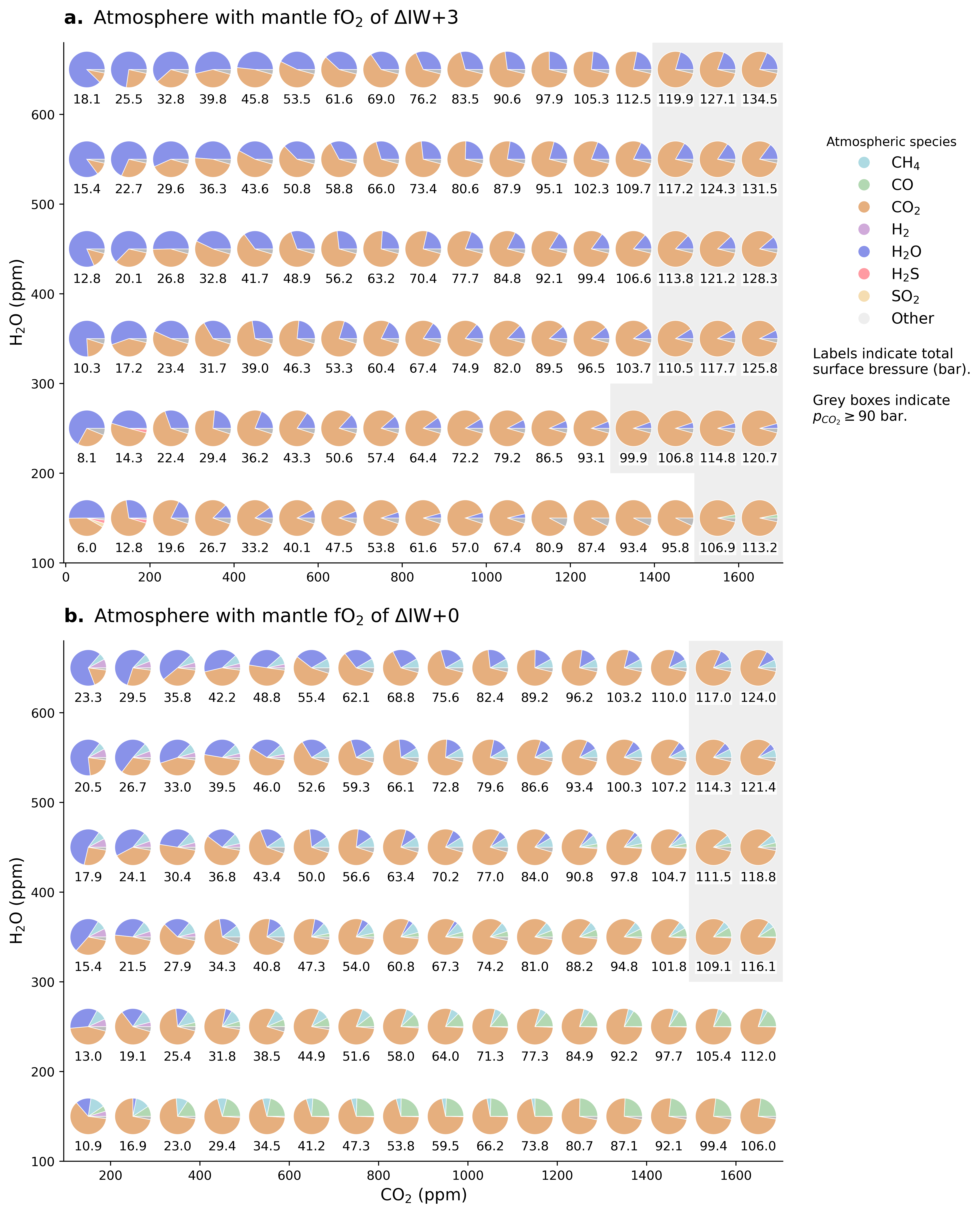}
    \caption{\textbf{Atmospheric composition as a function of mantle \ce{CO2} and \ce{H2O} content for two different mantle redox states}: \textbf{a.} an oxidised mantle with an oxygen fugacity ($f\ce{O2}$) of $\Delta$IW+3, and \textbf{b.} a more reduced mantle at $\Delta$IW+0. Each pie chart represents the molar fraction of species in the final atmosphere after 4~Gyr of volcanic degassing. Major species are shown in distinct colours; species with a molar fraction less than 3\% are grouped into the ``Other" category (grey). The labels beneath each pie indicate the total surface pressure in bar; grey shading indicates Venus-like conditions ($p\ce{CO2} \geq 90$~bar). }
    \label{fig:parameter_space}
\end{figure*}

The non-monotonic pressure in the water-poor scenario (Figure~\ref{fig:combined_composition}b) reflects the strongly $f$\ce{O2}-dependent solubility of carbon species in the melt. At $\Delta$IW $\le$ 0, carbon is speciated as CO and \ce{CH4} in the vapour, both of which are poorly soluble and partition efficiently into the gas phase, building a thick atmosphere. This is until $\Delta$IW $\approx$ +1, where the dominant carbon species shifts to \ce{CO2}, which is considerably more soluble in silicate melt; consequently, a greater fraction of the carbon budget is retained in the melt rather than degassing, and the atmosphere grows more slowly, producing the pressure dip  \citep{nicholls_convective_2025, ortenzi2020mantle}. Atmospheric growth resumes at higher $f$\ce{O2} driven by sulfur degassing \citep{nicholls_volatile_2026, o2021thermodynamic}.

These results suggest that on planets without volatile recycling, i.e., without an active form of tectonic subduction or other mechanisms of crustal recycling, and subsequent resurfacing of previously surfaced material, a massive secondary \ce{CO2} atmosphere like Venus’s is unlikely to form solely via continuous volcanic degassing (Figure~\ref{fig:combined_composition}a, b, c). This is consistent with \citet{ortenzi2020mantle}, whose results from using a 2-D thermal evolution model to simulate mantle convection and melt production also show that atmospheric pressures >40~bar are hard to achieve. In such settings, only a limited fraction of the mantle carbon inventory becomes available to the atmosphere before volatile exhaustion shuts down further growth.

To quantify the degree of volatile enrichment required to overcome the degassing bottleneck, we explore the sensitivity of atmospheric pressure to the initial mantle volatile inventory (Figure~\ref{fig:parameter_space}; complementary 2-D pressure maps are shown in Appendix Figure~\ref{fig:pressure_fO2}). In the absence of recycling, achieving Venus-like \ce{CO2} levels requires a mantle carbon content of at least 1450\,ppm of \ce{CO2}. Such strong enrichment suggests that, for planets with Earth-like bulk compositions, secondary volcanic degassing alone in the stagnant lid regime is insufficient to produce a secondary atmosphere rich in \ce{CO2}.

In contrast, in the recycling end-member (Figures~\ref{fig:combined_composition}d, e, f), no long-lived crustal reservoir is formed. Instead, after each degassing step, the residual extrusive melt is mixed back into the mantle. This limits progressive mantle depletion and allows carbon to be repeatedly remobilised by subsequent melting, enabling sustained degassing over the planet's lifetime. This end-member configuration produces substantially larger \ce{CO2} atmospheres.

Here, atmospheric \ce{CO2} pressures exceed 92~bar in highly oxidised, C-rich mantle scenarios. This demonstrates that volatile recycling, even in simplified or idealised form, is critical to achieving Venus-scale secondary atmospheres.

Beyond mantle composition and the presence of recycling, the final atmospheric state depends on the rate of magmatic delivery to the surface. In our model, this delivery is determined by the total melt production rate and the fraction of that melt that is extrusive (Equation~\ref{eqn:melt}). Because these two parameters are mathematically interchangeable in the calculation of erupted mass, we vary only the extrusive fraction as a proxy for both parameters, so the results are discussed for either scenario. Intrusive melts sequester their volatile load within the lithosphere (below the crust), whereas extrusive lavas reach the surface and facilitate direct degassing. (While intrusive melts may release some volatiles upon reaching saturation at depth, our model neglects this contribution.) To reflect the diversity of potential tectonic regimes, we test a range of extrusive fractions. Earth’s average extrusive fraction is estimated at 0.1--0.2 \citep{crisp1984rates}, with an upper limit of 0.6. For Venus, early Earth-like values suggest a ratio of $\sim$0.2 \citep{rozel2017continental}, though geo-dynamic models of crustal thickness and surface age suggest lower fractions are more probable \citep{lourenco2020plutonic, tian2023tectonics}.

\begin{figure}
    \centering
    \includegraphics[width=\columnwidth]{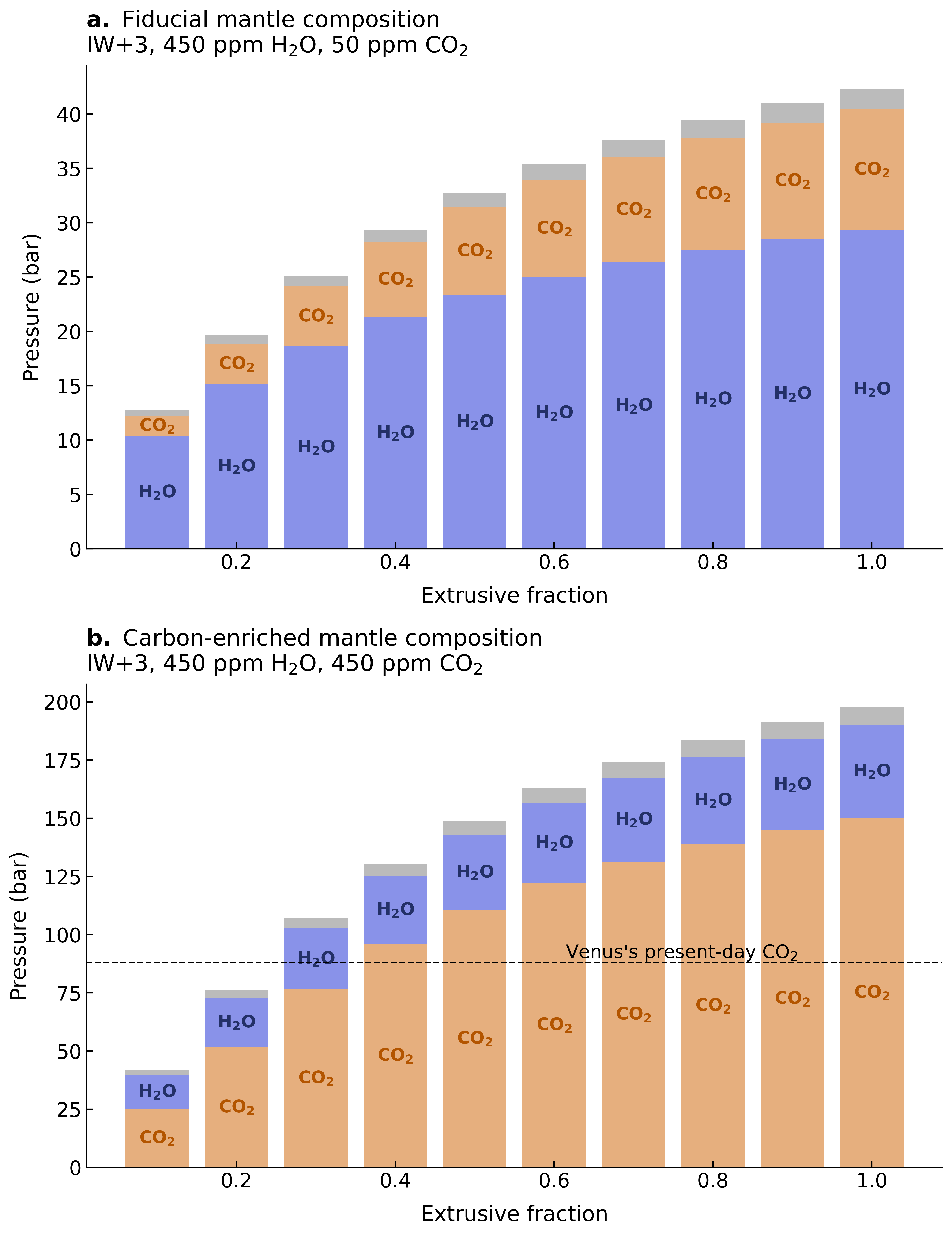}
    \caption{\textbf{Sensitivity of atmospheric composition to the volcanic extrusive-to-intrusive ratio.} Stacked bars show equilibrium partial pressures after 4\,Gyr of degassing as a function of the extrusive fraction (the ratio of extrusive to intrusive volcanism). All calculations are performed at a mantle oxygen fugacity of $\Delta$IW+3. \textbf{a,} Fiducial mantle composition (450\,ppm \ce{H2O}, 50\,ppm \ce{CO2}, 54\,ppm S and 1\,ppm N). \textbf{b,} Increased mantle \ce{CO2} content (450\,ppm). Cumulative bar heights represent total atmospheric pressure. Species contributing $<3\%$ of the total pressure are grouped in grey.}
    \label{fig:comp_extrusion_sweep}
\end{figure}

Figure~\ref{fig:comp_extrusion_sweep} shows the sensitivity of surface pressure to the extrusive fraction for the fiducial mantle composition, and a 450~ppm carbon-enhanced mantle. For the fiducial mantle case (adopting a fiducial mantle oxygen fugacity of $\Delta$IW+3, Figure~\ref{fig:comp_extrusion_sweep}a), the atmosphere remains well below Venus-like levels. In this regime, the total mantle carbon inventory acts as the primary bottleneck; even the most efficient magmatic delivery cannot overcome the limited volatile supply.

However, for the carbon-enriched mantle (450~ppm \ce{CO2}, Figure~\ref{fig:comp_extrusion_sweep}b), the extrusive fraction becomes the decisive factor in atmospheric evolution. While low extrusive fractions typical of modern Earth ($\sim$0.1--0.2) yield surface pressures between 45--75~bar, the 90~bar threshold of modern-day Venus is surpassed once the extrusive fraction exceeds $\sim$0.3. For highly extrusive scenarios where the ratio approaches 1.0, surface pressures can reach $\sim$200~bar. Across this extrusive-fraction sweep, the dominant effect is a scaling of total atmospheric pressure; atmospheric mixing ratios vary only weakly rather than remaining strictly constant. Thus, both the post-eruptive recycling of volatiles and the efficiency of their initial delivery to the surface are critical in determining the final atmospheric state.

\subsection{Crustal carbon accumulation}\label{sec:cca}
We next constrain the amount of \ce{CO2} that could have been sequestered into the crust during a hypothetical temperate phase on early Venus. Using a silicate weathering model for a habitable Venus, we calculate the maximum volcanic \ce{CO2} flux that can be matched by weathering without triggering atmospheric accumulation or runaway greenhouse conditions \citep{graham2020thermodynamic}.

\begin{figure}
    \centering
    \includegraphics[width=\columnwidth]{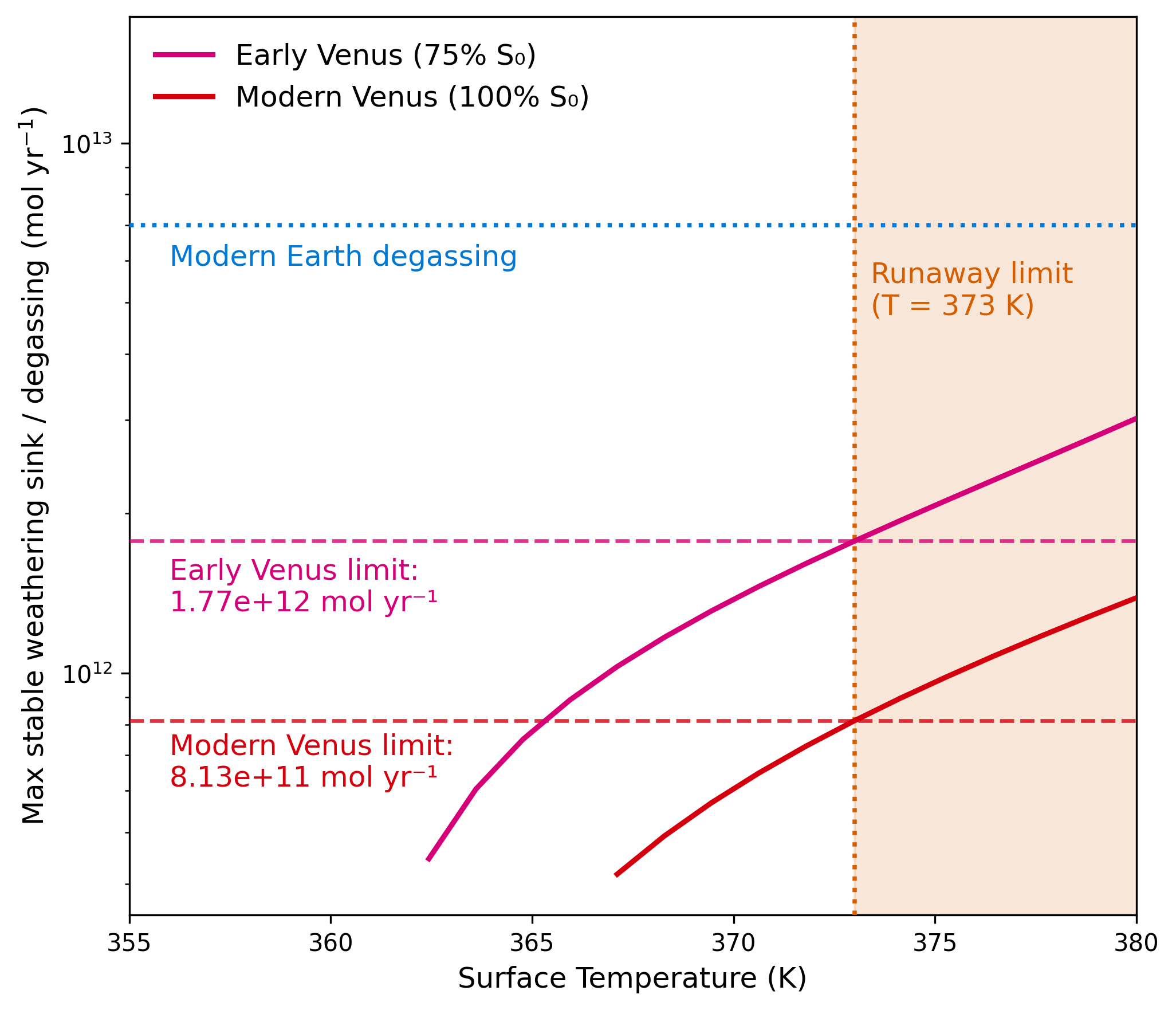}
    \caption{\textbf{Silicate weathering sink as a function of surface temperature under coupled climate--carbon equilibrium}, for early Venus receiving 75\% of present-day solar luminosity (pink) and Venus under present-day insolation (red). At each temperature, the atmospheric \ce{CO2} partial pressure is determined by radiative balance, and the weathering flux follows from the MAC model. The orange dotted vertical line marks our adopted liquid-water cut-off of $T = 373$~K, above which liquid water cannot persist at the surface. The orange shaded area is no longer physical, as the evaporation of oceans halts the weathering processes. Dotted horizontal lines indicate the maximum weathering flux each scenario can sustain before crossing this threshold ($1.77 \times 10^{12}$~mol~yr$^{-1}$ for early Venus; $8.13 \times 10^{11}$~mol~yr$^{-1}$ for modern Venus). The blue dashed line shows the modern Earth volcanic \ce{CO2} degassing rate for reference \citep{cynthia2019carbon, stewart2026igneous, coogan2026well}. In both Venus scenarios, the weathering sink at $T = 373$~K falls well below the modern Earth degassing rate, implying that silicate weathering could not have stabilised the climate against Earth-like volcanic outgassing once surface temperatures approached the adopted boiling point of water.}
    \label{fig:max_CO2}
\end{figure}

For an early Venus receiving 75\% of present-day solar luminosity, we find that steady climate--carbon equilibria exist for volcanic \ce{CO2} degassing rates up to $1.77 \times 10^{12}$~mol~yr$^{-1}$, as shown in Figure~\ref{fig:max_CO2}. Because we do not explicitly model the temporal evolution of the Sun, we also evaluate this threshold under present-day insolation ($8.13 \times 10^{11}$~mol~yr$^{-1}$) to bound the maximum influence that a warming Sun could have on our results. The threshold flux represents the weathering-limited upper bound on sustainable volcanic \ce{CO2} input: the largest degassing flux for which a habitable (<373~K) equilibrium climate state could exist. 

The factor of $\sim$4 gap between Venus's weathering capacity at the runaway threshold and Earth's volcanic flux \citep[$\sim7 \times 10^{12}$~mol~yr$^{-1}$;][]{cynthia2019carbon, stewart2026igneous, coogan2026well} reflects insolation alone: even early Venus absorbs $\sim$1.45 times more stellar flux than Earth, forcing radiative equilibrium to temperatures where energy-limited precipitation, suppressed carbonate kinetics, and near-zero p\ce{CO2} simultaneously throttle the weathering sink.

Assuming that Venus remained temperate for up to 30\% of its history \citep{warren2023narrow}, and that volcanic \ce{CO2} degassed is removed by silicate weathering, the maximum cumulative carbon sequestration would be $2.4 \times 10^{21}$~mol \ce{CO2} for early Venus conditions. This corresponds to $\sim$20~bar of \ce{CO2} when expressed as an equivalent surface pressure under hydrostatic equilibrium.

\begin{figure*}
    \centering
    \includegraphics[width=0.99\textwidth]{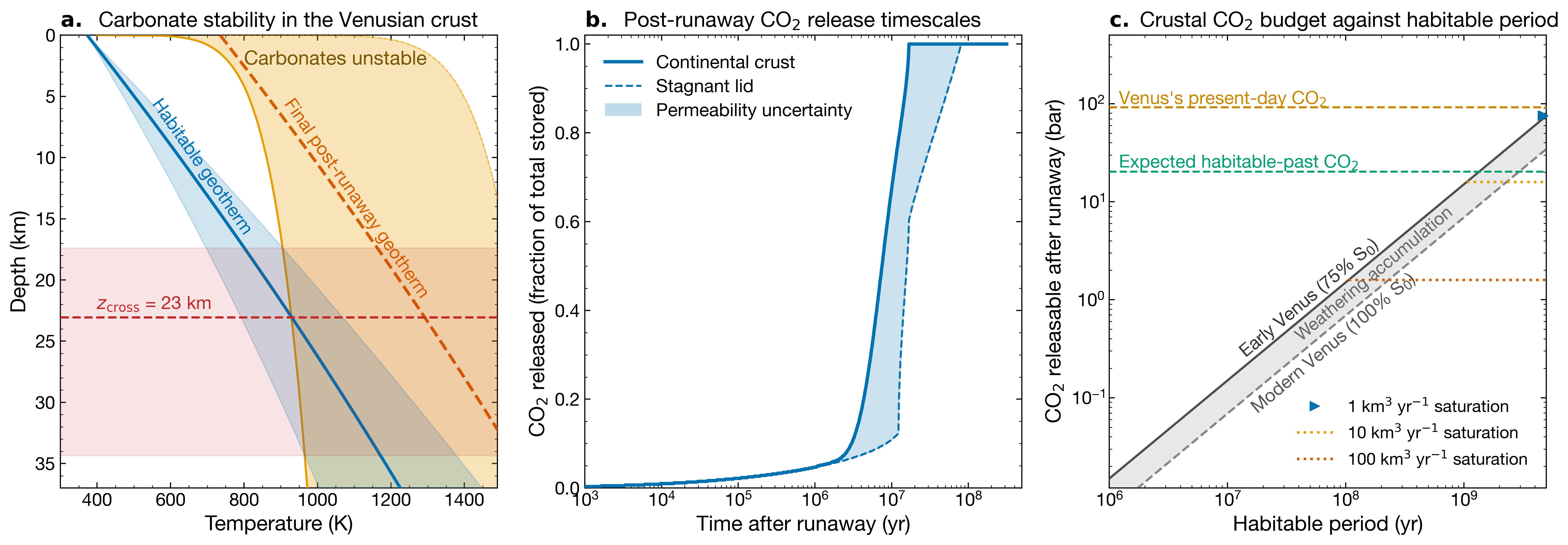}
    \caption{
    \textbf{Crustal carbonate stability and post-runaway \ce{CO2} release on
    a formerly temperate Venus.}
    \textbf{a.} Temperature--depth structure of the Venusian crust.
    The solid blue curve is the habitable-phase geotherm; blue shading shows the uncertainty for $q$. The dashed red curve is the post-runaway equilibrium geotherm ($T_s = 735$~K). The amber band is the carbonate instability field for nine decarbonation reactions \protect\citep{treiman2012mineral}; the left edge marks the lowest-energy reaction. The habitable geotherm intersects the instability field at the crossover depth $z_{\rm cross} = 23.1$~km (range $17.4$--$34.4$~km; crimson dashed line with uncertainty band): carbonates above this depth are stable during the habitable period, while the post-runaway geotherm destabilises the entire column.
    \textbf{b.} Cumulative fraction of the stored \ce{CO2} column released as a function of time after the runaway transition, for continental-crust permeability \protect\citep[solid blue;][]{manning1999permeability} and stagnant-lid permeability (dashed blue). Blue shading spans the permeability uncertainty. As shown in Appendix Figure~\protect\ref{fig:thermal_front}, thermal-front propagation (corrected for the latent heat of decarbonation) is the rate-limiting step; Darcy transit time is negligible ($<$1\% of the total).
    \textbf{c.} Total releasable \ce{CO2} versus habitable duration for resurfacing rates of 1, 10, and 100~km$^3$~yr$^{-1}$. The grey band shows weathering accumulation bounded by early- and modern-insolation cases. Dotted horizontal lines are the saturation ceilings for each resurfacing scenario; the right-edge arrow marks the 1~km$^3$~yr$^{-1}$ case, whose saturation time lies off-panel. The teal dashed line marks the expected habitable-past inventory (20.3~bar), and the gold dotted line marks present-day Venus atmospheric \ce{CO2} (92~bar).
    }
    \label{fig:decarb}
\end{figure*}

The depth structure of the carbonate stability field constrains which part of the crust contributes to this reservoir. As shown in Figure~\ref{fig:decarb}a, carbonates are thermally stable only above $z_{\rm cross} = 23$~km during the habitable phase; after the runaway transition the post-runaway geotherm destabilises the entire column, making the full stored inventory available for release.

The rate of release is governed by how rapidly heat diffuses downward. The cumulative fraction $f(t)$ released as a function of time after the transition is shown in Figure~\ref{fig:decarb}b. For continental-crust permeability, 50\% of the stored \ce{CO2} is released within $\sim$7.5~Myr, and 99\% within $\sim$17~Myr. Under stagnant-lid permeability, these timescales extend to $\sim$15.8~Myr and $\sim$78~Myr respectively. In both cases, the bulk of the stored carbon is remobilised on timescales short compared to Venus's inferred uninhabitable epoch of $\gtrsim$3~Gyr \citep{Way2020,Turbet2021, constantinou2024dry}, unless the habitable period was itself very brief.

How large the releasable inventory actually is depends on how long Venus was temperate, and on the volcanic resurfacing rate during that time. Figure~\ref{fig:decarb}c shows the total \ce{CO2} available for release as a function of habitable period $t_{\rm hab}$, for three resurfacing rates. During the habitable period, the crustal inventory grows linearly with time at the rate set by weathering. This linear growth continues until burial saturates the weathering sink at $t_{\rm bury}$ (Section~\ref{sec:c_release}), after which the inventory reaches a ceiling. For 100 and 10~km$^3$~yr$^{-1}$, saturation occurs at $\approx$106~Myr and $\approx$1.06~Gyr, with ceiling values of $\sim$1.6~bar and $\sim$15.9~bar respectively. At 1~km$^3$~yr$^{-1}$, saturation does not occur until $\sim$10.6~Gyr; the inventory therefore grows linearly throughout any plausible habitable period, reaching $\sim$75~bar at 5~Gyr. Matching Venus's present 92~bar inventory via this pathway alone would require a habitable period of $\sim$6.2~Gyr, exceeding the age of the Solar System.



For comparison, Figure~\ref{fig:primary_atmosphere} shows the surface pressures of primary atmospheres produced by magma ocean degassing, drawn from coupled interior-atmosphere planetary evolution models \citep{nicholls2024magma, nicholls2024magmadata}. Depending on redox state, C/H ratio, and total hydrogen inventory, primordial magma oceans can outgas hundreds of bars of \ce{CO2} within $\sim$1~Myr. Thus, while a secondary atmosphere can build up \ce{CO2} over Gyr via tectonic recycling and carbonate destabilisation, a primary atmosphere can generate equivalent surface pressures on much shorter timescales. 

When considering the net atmospheric pressure, we recognise that the primary \ce{H2O} is particularly susceptible to depletion via photodissociation and hydrogen escape \citep[e.g.,][]{lammer2006loss}. To this end, Figure~\ref{fig:primary_atmosphere} also isolates the pressure contribution from species excluding \ce{H2O}. This demonstrates that, even in the limit of total water loss, the degassing of non-condensable species during the magma ocean phase is sufficient to sustain surface pressures of \ce{CO2} in excess of $10^2$~bar.

\section{Discussion}\label{sec:discuss}
Three distinct mechanisms could, in principle, account for Venus's 92~bar \ce{CO2} inventory. First, efficient carbon outgassing during magma ocean crystallisation could directly produce a massive primary atmosphere, with the modern inventory representing whatever fraction survived early. Second, sustained secondary volcanic degassing into a persistently hot, dry surface environment could gradually accumulate atmospheric \ce{CO2} over Gyr timescales. Third, a formerly temperate Venus could have sequestered carbon into crustal carbonates via an active carbonate--silicate cycle, with its present atmosphere representing the remobilisation of that reservoir following a runaway greenhouse transition. Each pathway implies a fundamentally different climate and tectonic history. Our aim here is to quantify the atmospheric \ce{CO2} each can deliver, and assess whether any one of them (or their combination) can uniquely account for Venus's present inventory, and whether they are observationally distinguishable. We discuss the primary atmosphere scenario first, as it establishes the baseline against which secondary degassing and crustal remobilisation must be evaluated.

\subsection{Primary magma ocean outgassing}\label{sec:discuss_primary}
\begin{figure*}
\centering
\includegraphics[width=0.9\textwidth]{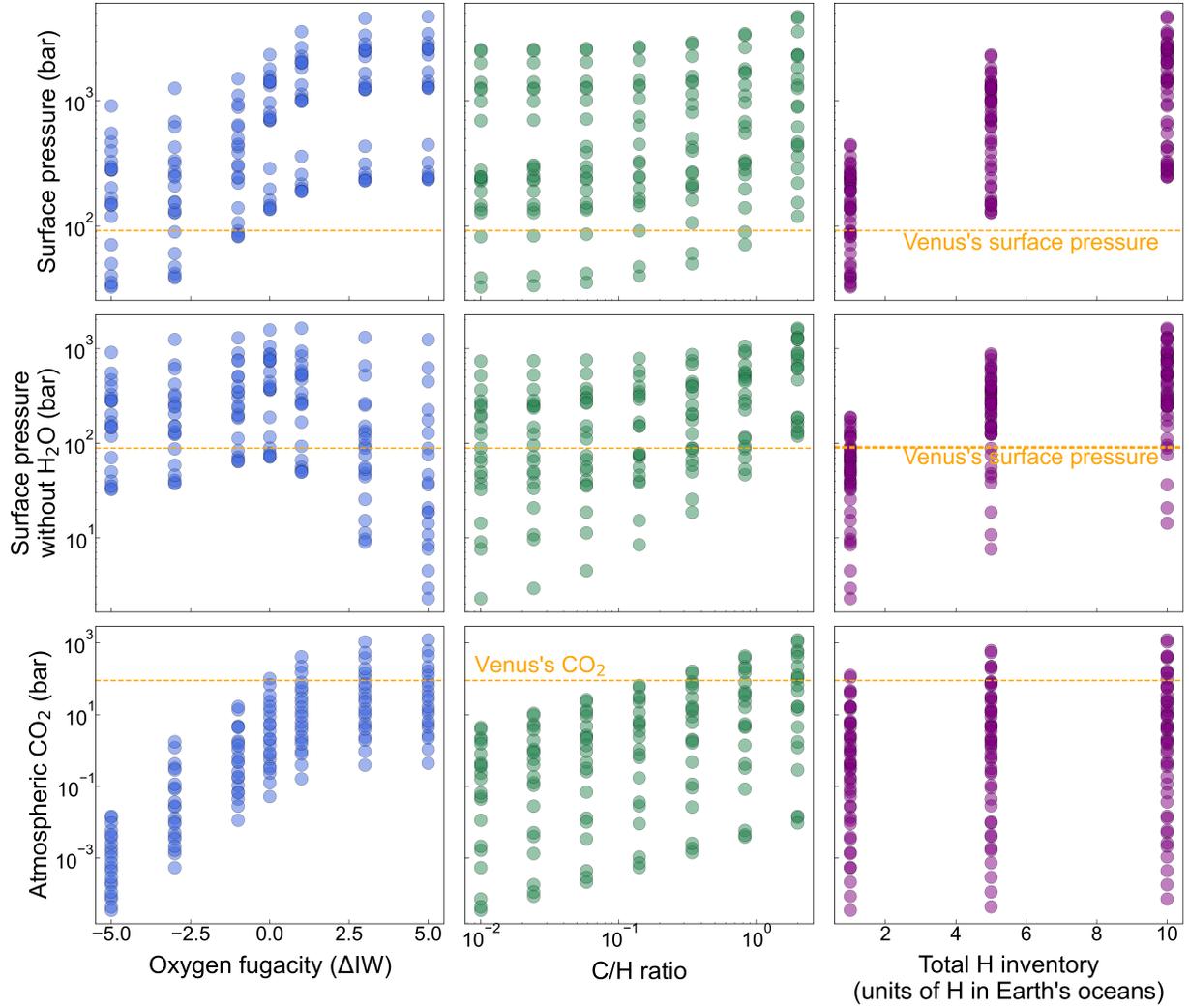}
\caption{\textbf{Primary atmospheres resulting from outgassing of early magma oceans under varying initial conditions}, from \protect\citet{nicholls2024magma, nicholls2024magmadata}. Top panels show the surface pressure, middle panel the surface pressure with \ce{H2O} removed, and bottom panels the atmospheric partial pressure of \ce{CO2}, each as a function of oxygen fugacity ($\Delta$IW, left), C/H ratio (middle), and total hydrogen inventory (right, given in units of Earth’s ocean mass).  Orange dashed lines indicate Venus’s present-day atmospheric pressure (top) and atmospheric \ce{CO2} (bottom) for reference. }
\label{fig:primary_atmosphere}
\end{figure*}

A primary atmosphere generated during magma ocean crystallisation can easily form a massive \ce{CO2} reservoir (Figure~\ref{fig:primary_atmosphere}; \citealt{nicholls2024magma}). The crux, however, is retaining this primary atmosphere. The late stages of accretion are dominated by giant impacts, which can strip significant fractions of a primitive atmosphere via bulk removal \citep[][and references therein]{schlichting2018atmosphere}. Stochastic impact simulations suggest that Earth could lose substantial portions of its primary envelope during the final assembly phase \citep{sinclair2020evolution, ghosh2025re}.

Even without invoking large stochastic impacts, XUV-driven hydrodynamic escape provides an efficient removal mechanism \citep{kulikov2006atmospheric, gillmann2009consistent}. Primordial envelopes, often rich in hydrogen from the protoplanetary nebula or water dissociation, are inflated and particularly susceptible to photo-evaporation \citep{owen2017evaporation}. While \ce{CO2} is heavier and more resistant to thermal escape than hydrogen, the extreme XUV flux from young stars drives vigorous hydrodynamic outflows that can entrain and drag heavier species, including C and O, into space \citep[][and references therein]{lammer2018origin}. However, recent 1D escape modelling shows that radiative cooling by carbon oxides and their photochemical products can suppress hydrodynamic escape in \ce{H2}-rich proto-atmospheres, reducing loss of heavier C-bearing species \citep{yoshida2024suppression}. While early atmospheric loss is plausible and supported by Venus’s noble gas record \citep{lammer2020constraining}, retention of a substantial primary \ce{CO2} reservoir remains a viable outcome.

If a significant fraction of the primary \ce{CO2} atmosphere is retained, the problem of explaining Venus's modern inventory is immediately solved, and the mantle carbon budget becomes largely irrelevant to the final  atmospheric state. It is only in the scenario where the primary atmosphere is  efficiently stripped that the planet becomes entirely dependent on its interior to reconstruct an atmosphere from a depleted mantle source. This harder scenario --- a planet that has lost its primary envelope and must rebuild through secondary degassing --- motivates the discussion that follows.

\subsection{Reproducing Venus's atmosphere with secondary degassing}\label{sec:discuss_secondary}
Our results identify a fundamental mass-balance constraint on the atmospheric evolution of stagnant lid planets. In the absence of crustal recycling, the mantle acts as a finite reservoir that is progressively depleted of its volatile inventory. The `one-way valve' of volatiles to the surface melt limits atmospheric \ce{CO2} accumulation to $\lesssim$25~bar (Figures~\ref{fig:combined_composition}a, b, c), even with a mantle \ce{CO2} nine times that of the fiducial case. 

To reach Venus's thick \ce{CO2} atmosphere through secondary degassing alone, the mantle source must be enriched with at least 1450~ppm \ce{CO2} (Figure~\ref{fig:parameter_space}). This concentration is thirty times greater than the 50~ppm estimated to remain in the silicate mantle after magma ocean solidification \citep{elkins2008linked}. In these solidification models, carbon partitions into a primary atmosphere, leaving the solid interior as a depleted reservoir comparable to Earth's mid-ocean ridge basalt source \citep[$\sim$30--110~ppm \ce{CO2};][and references therein]{hirschmann2009h, dasgupta2010deep}. While observations of modern Venusian degassing suggest that contemporary magmas may be carbon-rich \citep{constantinou2024dry}, the standard residual mantle inventory is insufficient to generate a 90~bar atmosphere under stagnant-lid conditions if the primary atmosphere is lost.

Indeed there are terrestrial precedents for high magmatic carbon enrichment. Enriched mantle sources on Earth can reach up to 1830~ppm \ce{CO2} \citep[oceanic island basalts;][]{hirschmann2009h}. On Earth, however, these highly enriched sources represent isolated pockets; our results require such enrichment to be characteristic of the entire volcanic source region to build a massive atmosphere. Achieving this could reflect either rapid magma-ocean solidification that traps carbon in the interior \citep{selsis2023cool,nicholls_convective_2025}, or long-lived deep volatile storage. In particular, a basal magma ocean has been proposed for Venus and could act as a persistent reservoir for incompatible components over Gyr timescales \citep{orourke2020venus}. Its relevance for atmospheric growth then depends on exchange with the convecting mantle: efficient coupling can resupply volcanic carbon, whereas strong isolation limits atmospheric expression.

Alternatively, Venus may have formed with a larger initial carbon budget. Current planetary formation models, which highlight extensive mixing among terrestrial protoplanets \citep{morbidelli2012building, izidoro2022origin}, suggest Venus and Earth should have accumulated similar volatile inventories. However, stochastic impacts \citep{sossi2022stochastic, dauphas2024bayesian} or heterogeneous sources \citep{broadley2022origin} can lead to substantially different starting inventories. While a significantly more \ce{CO2}-rich interior could build the atmosphere we see today, we have no other empirical evidence to support such a distinct starting composition for Venus. Even then, it is unlikely that the carbon will be in the mantle, as a large proportion would partition into the planet's core \citep{dasgupta2013ingassing}. 

Our findings provide a geochemical mechanism underpinning the geodynamic results of \citet{weller2023venus}. A simple stagnant lid regime is unlikely to generate a Venus-like \ce{CO2} atmosphere solely through secondary degassing under reasonable geochemical assumptions. Consequently, if Venus has operated under a growing stagnant lid for the majority of its history, its present atmosphere cannot be the product of steady-state mantle processing alone.

Our work, however, identifies processes which can overcome the mass-balance bottleneck in carbon degassing. The first is the total mass of magma delivered to the surface, which is determined by the melt production rate ($M_{\text{melt}}$) and the extrusive fraction ($r_{ei}$). As shown in Equation~\ref{eqn:melt}, these parameters are mathematically interchangeable in their effect on atmospheric growth. While we varied $r_{ei}$ to define our model limits, our findings equally describe scenarios where $r_{ei}$ remains Earth-like ($\sim$0.1--0.2) while the total melt production rate increases by the equivalent fraction.

By increasing the proportion of melt that reaches the surface to degas (rather than sequestering volatiles within intrusive magmas that stall in the lithosphere) the atmospheric build-up is accelerated.  To build a 90~bar \ce{CO2} atmosphere with a MORB-like mantle, the extrusive fraction must exceed $\sim$0.3 (Figure~\ref{fig:comp_extrusion_sweep}b). This requirement, however, introduces a geodynamic contradiction: low extrusive fractions ($\sim$0.1--0.2) are generally required to reconcile Venus's observed crustal thickness and tectonic structures \citep{lourenco2020plutonic, tian2023tectonics}. 

If $r_{ei}$ is indeed low, the 90~bar threshold must instead be reached through a higher total melt production rate. While modern magma production on Venus is low, estimated between 0.15 and 4~km$^3$\,yr$^{-1}$ \citep{nimmo1998volcanism, gillmann2014jgr, weller2021lpsc}, compared to the fiducial 27~km$^3$\,yr$^{-1}$, these rates likely reflect a period of quiescence. Melt production can increase by more than an order of magnitude during lithospheric overturns or tectonic transitions \citep{weller2021lpsc, weller2025sciadv,armann_simulating_2012}. Such high melt-flux regimes may be sustained by early giant impacts \citep{marchi2023natast} or self-regulating climate--interior feedbacks \citep{phillips2001grl}.

The second process to overcome the degassing bottleneck is suppression of long-lived crustal volatile storage. In EVolve, this is represented by the recycling end-member in which degassed extrusive residual melt is returned to the mantle, rather than accumulated in a crustal reservoir, thereby sustaining the mantle volatile supply available for later remelting and outgassing. On real planets, however, crustal recycling can still occur, even in stagnant-lid. In the limit of very high eruption rates (a heat-pipe-like end-member, as on Io), new crust is continuously emplaced and older crust is buried. Because crustal thickness cannot grow beyond the lithospheric thermal boundary layer, burial is accompanied by basal erosion, delamination and foundering, returning crustal material to the interior \citep[e.g.,][]{lourenco2020plutonic}. Our ``no-recycling'' and ``recycling'' configurations should therefore be interpreted as end-members of recycling efficiency, rather than literal tectonic states.

Whether recently proposed tectonic regimes for Venus, such as a `plutonic-squishy lid' \citep{lourenco2020plutonic} or a `deformable episodic lid' \citep{tian2023tectonics}, can provide recycling at the efficiency demanded by our results remains uncertain, given the absence of large-scale downwellings comparable to plate tectonics. These geodynamic mechanisms demonstrate that a ``born Venus" is physically plausible, but its existence requires a planetary history defined by high magmatic productivity, or efficient volatile transport.

Venus’s atmosphere is likely a hybrid of primary and secondary processes. We identify two possible scenarios. In the first, massive primary outgassing occurs, and the modern atmosphere represents the fraction of that primary \ce{CO2} that survived escape throughout Venus's lifetime. In this case, the mantle is depleted early, and subsequent volcanism contributes little to the total atmospheric inventory. In the second, the crystallisation dynamics of a short-lived magma ocean limit primary outgassing, favouring the retention of carbon in the planet's deep interior \citep{hier_volatile_2017,sim_volatile_2024}. This stored carbon later builds a massive atmosphere through secondary degassing. While the relative contributions of these processes remain uncertain, both scenarios converge on the same geochemical prerequisite: a carbon-rich mantle.

We now turn to whether a habitable past offers a third, distinct pathway to the same present-day atmospheric inventory.

\subsection{Considering a habitable past}\label{sec:discuss_hab}
An alternative explanation for Venus’s present atmosphere is that the planet once sustained surface liquid water and an active carbonate–silicate cycle, with its current \ce{CO2}-rich atmosphere resulting from the subsequent thermal destabilisation of crustal carbonates following climate collapse. 

Within this scenario, the coupled climate–weathering framework presented here places an upper bound on the amount of carbon which could have been stored in the crust prior to a runaway greenhouse transition. This upper bound is constrained by finding the limit at which sequestration can keep up with volcanic degassing, under a temperate climate. Under this regime, the resulting crustal reservoir contains only $\sim$20~bar of \ce{CO2} (Figure~\ref{fig:decarb}c). This inventory is roughly one-fifth of the 92~bar observed today. The destabilisation of a carbonate reservoir sequestered under strictly temperate conditions is thus insufficient, by itself, to account for Venus’s present atmosphere.

For context, the maximum weathering-limited sequestration rate ($1.77 \times 10^{12}$~mol~yr$^{-1}$) is an order of magnitude larger than the volcanic \ce{CO2} supply in the fiducial secondary degassing scenario ($1.14 \times 10^{11}$~mol~yr$^{-1}$), and comparable only to the carbon-enriched case ($1.02 \times 10^{12}$~mol~yr$^{-1}$). In other words, for a temperate Venus to accumulate its maximum crustal reservoir, it would require a volcanic supply close to the upper end of our secondary degassing models --- the same conditions that, in an uninhabitable setting, are needed to build a 90~bar atmosphere directly.

As shown in Figure~\ref{fig:decarb}b, the stored carbonate column is fully released on timescales of 17--78~Myr after the transition --- short compared to Venus's  uninhabitable epoch of $\gtrsim$3~Gyr \citep{Way2020, Turbet2021, constantinou2024dry}, implying that any carbonate reservoir formed during a temperate phase would have been fully degassed long before the present day. 

The shortfall between the maximum sequestered inventory of $\sim$20~bar and the observed 92~bar suggests that a habitable past alone cannot explain the present atmosphere. This 20~bar limit could be too conservative an estimate; omitted climate feedbacks like cloud--albedo effects, or a higher planetary albedo could allow temperate conditions to persist at higher degassing rates, increasing both carbonate sequestration and post-runaway released \ce{CO2} \citep{Way2016, charnay_cloud_2017, Turbet2021, Way2020, goldblatt_cloud_2021}. Otherwise, the remainder must originate from another source --- either primary atmosphere being retained, secondary volcanic degassing as explored in Section~\ref{sec:secondary}, or some combination of the two. 

Ultimately, a past runaway greenhouse scenario is insufficient on its own, and is no more consistent with the observed atmosphere than an always-hot evolutionary pathway.

\subsection{Distinguishing dead Earths from Venuses}\label{sec:discuss_exo}

Identifying former habitability on rocky exoplanets is hindered by the potential degeneracy of atmospheric states. Our results show that atmospheric \ce{CO2} alone is not a sufficient diagnostic to distinguish a planet that was once temperate from one that was always hot. A massive \ce{CO2} atmosphere can arise from at least three distinct evolutionary pathways, each implying a different history for the planet's climate and tectonics.

The first pathway is the retention of a primary atmosphere. If a carbon-rich, oxidised magma ocean produces a massive primary envelope that survives early stellar activity and impact erosion, the planet will begin its life with a dense atmosphere. The second pathway is the slow growth of a secondary atmosphere. As we have shown, continuous volcanic outgassing can eventually build a Venus-like atmosphere, provided the interior is carbon-enriched and the planet possesses either a high rate of extrusion, or a mechanism for crustal recycling. The third pathway is the catastrophic remobilisation of a sequestered carbon reservoir. In this ``dead Earth" scenario, a formerly temperate planet destabilises and releases its crustal carbonates following a runaway greenhouse event or some stochastic heating event.

Because these different trajectories can converge on the same final atmospheric state, the presence of a 90~bar \ce{CO2} envelope is an ambiguous signal. To break the degeneracy between a planet that was ``born" hot or ``died" later, and to map the true boundaries of the habitable zone, we will likely need to move beyond relying on \ce{CO2} mass alone, and look for additional chemical, isotopic, or even geographical tracers of a planet's volcanic and climatic past.
 
\section{Conclusions}\label{sec:concl}

Our results show that a massive \ce{CO2} atmosphere does not prescribe a unique evolutionary history. Instead, atmospheric mass is a degenerate signal that fails to distinguish between a planet that was born hot and one that underwent a climatic catastrophe. 

A past habitable climate provides one pathway to a massive \ce{CO2} atmosphere. A temperate world with an active carbonate-silicate cycle could sequester up to $\sim$20~bar equivalent \ce{CO2} in its crust. Following a transition to a runaway greenhouse, heat diffusion liberates the stored carbon within 17--78~Myr. This timescale is short compared to Venus's inferred uninhabitable epoch of $\gtrsim$3~Gyr, so the reservoir would be fully degassed long before the present day. In this scenario, a past runaway greenhouse is on its own insufficient to explain Venus's present \ce{CO2}-rich atmosphere. 

Sustained secondary degassing provides an alternative pathway to a massive \ce{CO2} atmosphere. A stagnant-lid planet outgassing an Earth-like mantle is limited to $\sim$25~bar. This mass-balance constraint is bypassed if the tectonic regime facilitates crustal recycling, or if the magmatic flux to the surface is increased through either higher extrusive fractions or higher total melt production. Under these conditions, or in the presence of a highly carbon-enriched mantle, atmospheric \ce{CO2} can readily exceed 100~bar.  Venus's modern atmosphere could thus plausibly be the product of secondary degassing. If even a fraction of the primary atmosphere is also retained, these pressures can be higher still. Early magma ocean degassing can deliver hundreds of bars of \ce{CO2} within $\sim$1~Myr, although the fraction of \ce{CO2} retained after early hydrodynamic losses and longer-term atmospheric erosion remains a key uncertainty.

Because these three distinct trajectories can produce overlapping atmospheric \ce{CO2} inventories under certain conditions, the presence of a dense \ce{CO2} envelope is an ambiguous signal. This ambiguity is not universal: a habitable-past scenario is, on its own, incapable of reproducing Venus's full 92~bar inventory, whereas secondary degassing under favourable conditions can. However, the overlap is subject to the mantle physics and geochemistry, which are observationally inaccessible, or degenerate. Thus, atmospheric \ce{CO2} alone is insufficient as a reliable diagnostic of planetary history. Breaking this possible equifinality will require identifying alternative tracers that can distinguish the remains of a formerly temperate world from a planet that was never habitable.

\section*{Acknowledgements}
T.C. thanks the Science and Technology Facilities Council (STFC) for the PhD studentship (grant reference ST/X508299/1). OS acknowledges funding from STFC grant UKRI1184.  T.C. thanks Josh Shea for discussions on carbon within the mantle. 

\section*{Data Availability}
No new data were used by or were generated for this work.
The EVolve model used here, including the versions of both EVo (the submodel handles volcanic degassing) and FastChem used to produce our results, is described in \citet{liggins2022growth}, and is freely available on GitHub at \url{https://github.com/pipliggins/EVolve}. The FastChem 2.0 model of \citet{stock2022fastchem} is also freely available. Primary atmosphere model outputs drawn from \citet{ nicholls2024magmadata}.



\bibliographystyle{mnras}
\bibliography{equifinality} 




\appendix

\section{Outgoing Longwave Radiation (OLR) Parameterisation}\label{appendix:OLR}

The Outgoing Longwave Radiation (OLR) is calculated as a function of surface temperature ($T_s$) and the partial pressure of \ce{CO2} ($p\text{CO}_2$) using the polynomial fit derived from radiative-convective models by \citet{graham2020thermodynamic}. The OLR flux (in W~m$^{-2}$) is given by:

\begin{equation}
    \text{OLR}(T_s, p\text{CO}_2) = I_0 + \sum_{i=0}^6 \sum_{j=0}^4 k_{ij} \cdot \xi_i(T_s) \cdot \nu_j(p\text{CO}_2)
\end{equation}

where $I_0 = -3.1$~W~m$^{-2}$ is a constant offset. The basis functions for temperature ($\xi$) and pressure ($\nu$) are defined as:

\begin{align}
    \xi_n(T_s) &= \left[ 0.01 \cdot (T_s - 250) \right]^n \\
    \nu_m(p\text{CO}_2) &= 
    \begin{cases} 
      \left[ 0.2 \cdot \log_{10}(p\text{CO}_2) \right]^m & \text{if } p\text{CO}_2 < 1 \text{ bar} \\
      \left[ \log_{10}(p\text{CO}_2) \right]^m & \text{if } p\text{CO}_2 \ge 1 \text{ bar}
   \end{cases}
\end{align}

The coefficient matrices $k_{ij}$ differ between the low-pressure ($<1$ bar) and high-pressure ($\ge 1$ bar) regimes. The specific values used in this study are provided in Tables \ref{tab:olr_low} and \ref{tab:olr_high}.

\begin{table}
\centering
\caption{OLR Coefficients ($k_{ij}$) for Low Pressure Regime ($p\text{CO}_2 < 1$ bar)}
\label{tab:olr_low}
\begin{tabular}{c c c c c c}
\hline
$i \backslash j$ & $\nu_0$ & $\nu_1$ & $\nu_2$ & $\nu_3$ & $\nu_4$ \\
\hline
$\xi_0$ & 87.8373 & -311.289 & -504.408 & -422.929 & -134.611 \\
$\xi_1$ & 54.9102 & -677.741 & -1440.63 & -1467.04 & -543.371 \\
$\xi_2$ & 24.7875 & 31.3614 & -364.617 & -747.352 & -395.401 \\
$\xi_3$ & 75.8917 & 816.426 & 1565.03 & 1453.73 & 476.475 \\
$\xi_4$ & 43.0076 & 339.957 & 996.723 & 1361.41 & 612.967 \\
$\xi_5$ & -31.4994 & -261.362 & -395.106 & -261.600 & -36.6589 \\
$\xi_6$ & -28.8846 & -174.942 & -378.436 & -445.878 & -178.948 \\
\hline
\end{tabular}
\end{table}

\begin{table}
\centering
\caption{OLR Coefficients ($k_{ij}$) for High Pressure Regime ($p\text{CO}_2 \ge 1$ bar)}
\label{tab:olr_high}
\begin{tabular}{c c c c c c}
\hline
$i \backslash j$ & $\nu_0$ & $\nu_1$ & $\nu_2$ & $\nu_3$ & $\nu_4$ \\
\hline
$\xi_0$ & 87.8373 & -52.1056 & 35.2800 & -1.64935 & -3.42858 \\
$\xi_1$ & 54.9102 & -49.6404 & -93.8576 & 130.671 & -41.1725 \\
$\xi_2$ & 24.7875 & 94.7348 & -252.996 & 171.685 & -34.7665 \\
$\xi_3$ & 75.8917 & -180.679 & 385.989 & -344.020 & 101.455 \\
$\xi_4$ & 43.0076 & -327.589 & 523.212 & -351.086 & 81.0478 \\
$\xi_5$ & -31.4994 & 235.321 & -462.453 & 346.483 & -90.0657 \\
$\xi_6$ & -28.8846 & 284.233 & -469.600 & 311.854 & -72.4874 \\
\hline
\end{tabular}
\end{table}

\begin{figure*}
    \centering
    \includegraphics[width=0.9\textwidth]{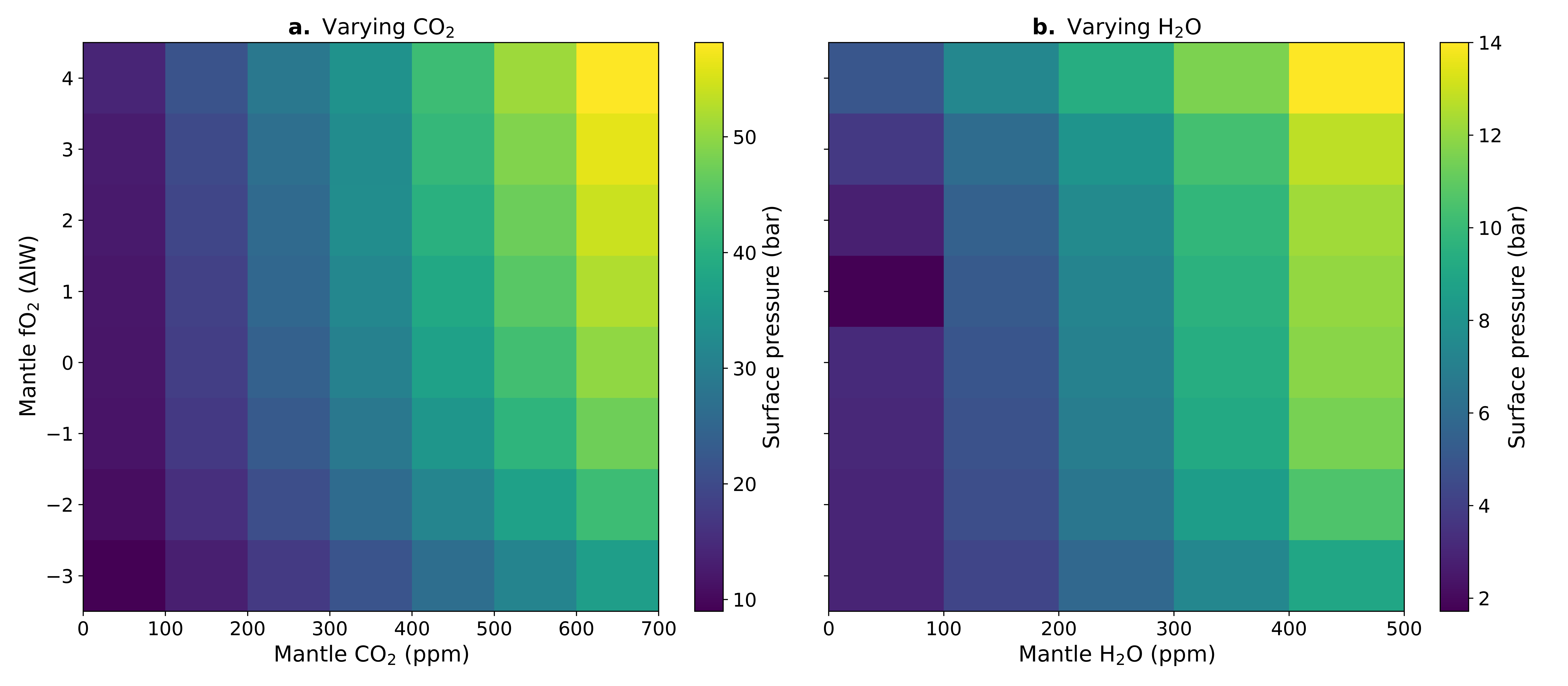}
    \caption{Final surface atmospheric pressure after 4~Gyr of volcanic degassing, shown as a function of mantle oxygen fugacity (fO2) and \textbf{a.} initial mantle \ce{CO2} content, \textbf{b.} initial mantle \ce{H2O} content. Other mantle volatiles are held constant at fiducial values: 450~ppm \ce{H2O} \protect\citep{elkins2008linked}, 50~ppm \ce{CO2} \protect\citep{elkins2008linked}, 54~ppm S, and 1~ppm N (scaled to the \ce{CO2} content of Earth's depleted MORB; \protect\citet{ding2017fate, marty2003nitrogen, le2017heterogeneity} respectively).}
    \label{fig:pressure_fO2}
\end{figure*}

\begin{figure}
    \centering
    \includegraphics[width=\columnwidth]{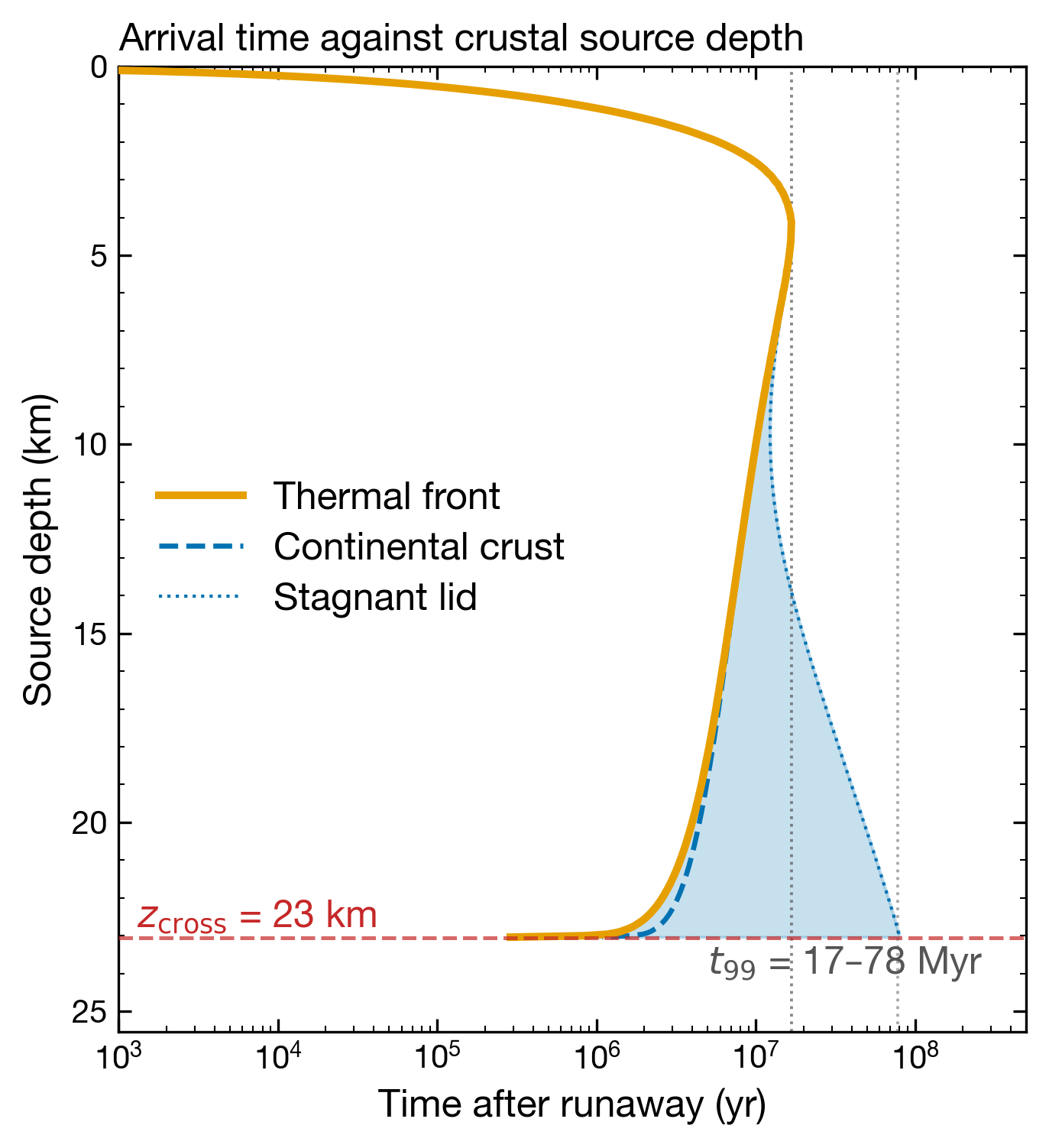}
    \caption{\textbf{Thermal-front and \ce{CO2} arrival time as a function of crustal source depth, for the post-runaway decarbonation model. }The orange curve shows the thermal-front arrival time. The blue dashed and dotted curves show the total \ce{CO2} arrival time for the continental crust \citep{manning1999permeability} and stagnant-lid ($k \times 10^{-2}$) permeability end-members respectively; the shaded band spans the permeability uncertainty. The near-coincidence of the orange and blue curves at all depths confirms that Darcy transit time contributes less than 1\% of the total, and that heat diffusion is the sole rate-limiting step. The red dashed line marks $z_{\rm cross} = 23$~km, the base of the stable carbonate column. The grey dotted vertical lines bracket $t_{99} = 17$--$78$~Myr, the time by which 99\% of stored \ce{CO2} has reached the surface across the full permeability range. The $y$-axis is inverted to show depth increasing downward.}
    \label{fig:thermal_front}
\end{figure}

\section{Carbon Inventory Bookkeeping and Pressure Conversion}\label{appendix:bookkeeping}

For the crustal-sequestration calculation, the cumulative sequestered \ce{CO2} inventory is
\begin{equation}
n_{\rm seq}(t_{\rm hab}) = F_w\,\min\!\left(t_{\rm hab}, t_{\rm bury}\right),
\end{equation}
where $F_w$ is the weathering-limited sink (mol~yr$^{-1}$), $t_{\rm hab}$ is habitable duration, and
\begin{equation}
t_{\rm bury}=\frac{z_{\rm cross}}{v}, \qquad v=\frac{\dot V}{A_{\rm planet}}.
\end{equation}

The corresponding equivalent surface pressure is
\begin{equation}
P_{\rm seq}=\frac{n_{\rm seq} M_{\rm CO_2} g}{A_{\rm planet}}.
\end{equation}

Post-runaway atmospheric delivery is computed as
\begin{equation}
P_{\rm rel}(t)=f(t)\,P_{\rm seq},
\end{equation}
where $f(t)\in[0,1]$ is the cumulative released fraction from the thermal-front/Darcy model.

For convenience, the molar inventory equivalent to 1~bar is
\begin{equation}
n_{1\rm bar}=\frac{10^5 A_{\rm planet}}{M_{\rm CO_2} g},
\end{equation}
so pressures can be written as $P=n/n_{1\rm bar}$ (bar).

\section{Mineral Decarbonation Equilibria}\label{appendix:carbonates}

We evaluate the stability of crustal carbonates against thermal metamorphism using nine key decarbonation equilibria derived from \citet{treiman2012mineral}. The equilibrium \ce{CO2} pressure ($P_{\text{CO}_2}$) for each reaction is determined as a function of temperature ($T$) using the enthalpy of reaction ($\Delta H_{\text{rxn}}$)
\begin{equation}
    \ln P_{\text{CO}_2} \approx -\frac{\Delta H_{\text{rxn}}(T)}{RT} + C,
\end{equation}
where $R$ is the ideal gas constant (8.314 J mol$^{-1}$ K$^{-1}$) and $C = \Delta S_{\rm rxn}/R$ is set by the standard entropy change of the reaction. Because $\Delta H_{\text{rxn}}$ varies with temperature, we interpolate values linearly between the reference temperatures provided in Table \ref{tab:decarb_reactions} to calculate the stability curves shown in the main text. 

At depths where the geotherm temperature exceeds the tabulated range, $\Delta H_{\rm rxn}$ values are linearly extrapolated; the weak temperature dependence of $\Delta H$ ($<2$~kJ~mol$^{-1}$ per 100~K) means this introduces negligible error.

\begin{table*}[t]
\centering
\begin{minipage}{\textwidth}
    \centering
    \caption{Enthalpies of reaction ($\Delta H_{\text{rxn}}$) for carbonate-silicate equilibria used to model crustal decarbonation. Values are in kJ mol$^{-1}$ at specified temperatures. Data sourced from \citet{treiman2012mineral}.}
    \label{tab:decarb_reactions}
    \renewcommand{\arraystretch}{1.2}
    \begin{tabular}{l c c c c}
    \hline
    \textbf{Equilibrium Reaction} & \multicolumn{4}{c}{\textbf{$\Delta H_{\text{rxn}}$ (kJ mol$^{-1}$)}} \\
     & \textbf{600 K} & \textbf{700 K} & \textbf{800 K} & \textbf{900 K} \\
    \hline
    1. \ce{CaCO3 + MgSiO3 + SiO2 <-> CaMgSi2O6 + CO2} & 65.1 & 63.3 & 61.3 & 58.7 \\
    2. \ce{SiO2 + MgCO3 <-> MgSiO3 + CO2} & 82.8 & 81.4 & 79.4 & 76.9 \\
    3. \ce{CaCO3 + SiO2 <-> CaSiO3 + CO2} & 86.5 & 84.9 & 83.0 & 80.6 \\
    4. \ce{CaMg(CO3)2 + SiO2 <-> CaCO3 + MgSiO3 + CO2} & 87.5 & 86.3 & 84.4 & 82.1 \\
    5. \ce{MgSiO3 + MgCO3 <-> Mg2SiO4 + CO2} & 89.0 & 87.1 & 85.0 & 82.6 \\
    6. \ce{CaMg(CO3)2 + MgSiO3 <-> CaCO3 + Mg2SiO4 + CO2} & 93.7 & 92.0 & 90.0 & 87.8 \\
    7. \ce{MgCO3 <-> MgO + CO2} & 116.8 & 115.7 & 114.3 & 110.5 \\
    8. \ce{CaMg(CO3)2 <-> CaCO3 + MgO + CO2} & 120.4 & 119.2 & 117.6 & 115.7 \\
    9. \ce{CaCO3 <-> CaO + CO2} & 176.0 & 174.6 & 173.4 & 171.9 \\
    \hline
    \end{tabular}
\end{minipage}
\end{table*}


\bsp	
\label{lastpage}
\end{document}